\documentclass[prc,english,twocolumn,showpacs,superscriptaddress,floatfix]{revtex4-1}
\usepackage{babel}
\usepackage{ucs} 
\usepackage[utf8x]{inputenc}
\usepackage{amsmath}
\usepackage{amssymb}
\usepackage{amsfonts}
\usepackage[dvips]{graphicx}
\usepackage{verbatim}
\usepackage{subfigure}
\usepackage{numprint}
\usepackage{multirow}

\newcommand{\ket}[1]
	{|#1\rangle}
\newcommand{\bra}[1]
	{\langle #1 |}

\newcommand{\nn}{\notag \\}

\newcommand{\op}[1]
    {\mathrm{\hat{#1}}}

\newcommand{\barh}{\mathrm{\bar{H}}}
\newcommand{\nmax}{N_\textrm{max}}
\newcommand{\nlo}{N${^3}$LO}

\newsavebox{\fmbox}

\newcommand{\SE}{Schrödinger equation}

\newcommand{\WE}{Wigner-Eckart theorem}
\newcommand{\CG}{Clebsch-Gordon coefficient}

\begin{document}
\title{Spherical coupled-cluster theory for open-shell nuclei}
\date{\today}

\author{G.~R.~Jansen}
\email{gustav.jansen@utk.edu}

\affiliation{Department of Physics and Astronomy, University of
    Tennessee, Knoxville, Tennessee 37996, USA}
\affiliation{Physics Division, Oak Ridge National Laboratory, Oak
  Ridge, Tennessee 37831, USA}
\affiliation{Department of Physics and Center of Mathematics for
  Applications, University of Oslo, N-0316 Oslo, Norway}


\begin{abstract}
\begin{description}
\item[Background] A microscopic description of nuclei is important to understand
    the nuclear shell-model from fundamental principles. This is difficult to
    achieve for more than the lightest nuclei without an effective approximation scheme.
\item[Purpose] Define and evaluate an approximation scheme that can be used to study nuclei
    that are described as two particles attached to a closed (sub-)shell
    nucleus.
\item[Methods] The equation-of-motion coupled-cluster formalism has been used to
    obtain ground and excited state energies. This method is based on the
    diagonalization of a non-Hermitian matrix obtained from a similarity
    transformation of the many-body nuclear Hamiltonian. A chiral interaction at
    the next-to-next-to-next-to leading order~(\nlo) using a cutoff at $500$~MeV
    was used.
\item[Results] The ground state energies of
    ${}^6$Li and ${}^6$He were in good agreement with a no-core
    shell-model calculation using the same interaction. Several excited states were also produced with
    overall good agreement. Only the $J^\pi=3^+$
    excited state in ${}^6$Li showed a sizable deviation. The
    ground state energies of ${}^{18}$O, ${}^{18}$F and ${}^{18}$Ne were converged,
    but underbound compared to experiment.
    Moreover, the calculated spectra were converged and comparable to both experiment and shell-model studies
    in this region. Some excited states in ${}^{18}$O were high or missing in
    the spectrum. It was also shown that the wave function for both ground and
    excited states separates into an intrinsic part and a Gaussian for the
    center-of-mass coordinate. Spurious center-of-mass excitations are clearly identified.
\item[Conclusions] Results are converged with respect to the size of the model
    space and the method
    can be used to describe nuclear states with simple structure. Especially the
    ground state energies were very close to what has been achieved by exact diagonalization.
    To obtain a closer match with experimental data, effects of
    three-nucleon forces, the scattering continuum as well as additional
    configurations in the coupled-cluster approximations, are necessary.
\end{description}

\end{abstract}

\maketitle

\section{Introduction}
In the past decade, the computing resources made available for
scientific research has grown several orders of magnitude. This trend will
continue during this decade, culminating in exascale computing facilities. 
This will promote new insights in every discipline as new problems can
be solved and old problems can be solved faster and to a higher precision. 

In nuclear physics, one important goal is a predictive theory, where
nuclear observables can be calculated from first principles. But even with the
next generation of supercomputers, a virtually exact solution to the nuclear many-body problem is possible only for light
nuclei~(see \textcite{Leidemann2013} for a recent review on many-body methods).
Using a finite basis expansion, a full diagonalization can currently be performed for
nuclei in the $p$-shell region~\cite{Navratil2009}. This might be extended to
light $sd$-shell nuclei within the next couple of years with access to sufficient computing resources. For
\emph{ab initio} access to larger nuclei, the problem has to be approached
differently~\cite{Dickhoff2004,Hagen2007b,Roth2007,Tsukiyama2011,Soma2013}.

In coupled-cluster theory, a series of controlled approximations are performed
to generate a similarity transformation of the nuclear Hamiltonian. At a given
level of approximation, efficient formulas exist to evaluate the ground state
energy of a closed (sub-)shell reference nucleus. The similarity-transformed
Hamiltonian is then diagonalized to calculate excited states and states of
nuclei with one or more valence nucleons. This defines the equation-of-motion
coupled-cluster (EOM-CC) framework (see
\textcite{Bartlett2007} for a recent review and  \textcite{shavittbartlett} for a
textbook presentation).
Recently, this method was applied to the oxygen~\cite{Hagen2012a} and the
calcium~\cite{Hagen2012b} isotopic chains, as well as
${}^{56}$Ni~\cite{Binder2012}, extending the reach of \emph{ab initio}
methods in the medium mass region. Further calculations in the nickel and tin regions
are also planned.

In this work, I will refine the EOM-CC method
for two valence nucleons attached to a closed (sub-)shell reference (2PA-EOM-CC). The
general theory was presented in \textcite{Jansen2011}, where calculations were
limited to small model spaces.
The working equations are now completely
reworked in a spherical formalism. Since the Hamiltonian is invariant under
rotation, this formalism enables us to do calculations in significantly larger model
spaces. All relevant equations of spherical formalism are explicitly included here for future reference.
The method, in the form presented here, has already been successfully applied to
several nuclei~\cite{Hagen2012a,Hagen2012b, Ekstrom2013,
Lepailleur2013}, but the formalism has not been presented.

A brief overview of general coupled-cluster theory and the equation-of-motion
extensions is given in Sec.~\ref{sec:theory}.  In Sec.~\ref{sec:2pa} I
derive the working equations for 2PA-EOM-CC and discuss numerical results for
selected $p$- and $sd$-shell nuclei in Sec.~\ref{sec:results}. As a proper
treatment of three body forces and continuum degrees of freedom is beyond the
scope of this article, the focus will be on convergence, rather than comparison
to experiment.
Finally, in Sec.~\ref{sec:conclusion} I present conclusions and discuss
the road ahead. All angular momentum
transformations used in this work are defined in the appendix.

\section{\label{sec:theory}Coupled-cluster theory}
In this section the Hamiltonian that enters the coupled-cluster calculations is
defined. I have also included a brief review of single reference coupled
cluster-theory together with the equation-of-motion (EOM-CC) extensions. In this
framework, a diagonalization in a truncated vector space  yields excited states,
where also nuclei with different particle numbers can be approached by
choosing an appropriate basis. 
The presentation is kept short and is focused on the aspects important
for deriving the spherical version of the 2PA-EOM-CCSD method presented in
Sec.~\ref{sec:2pa}.

All calculations are done using the intrinsic Hamiltonian
\begin{equation}
    \op{H} = \left( 1 - \frac{1}{A^*} \right) \sum_{i=1}^A \frac{p_i^2}{2m} + \left[\sum_{i<j=1}^A \op{v}_{ij} -\frac{\vec{p}_i \cdot \vec{p}_j}{mA^*}\right] \ . 
    \label{eq:hamilton}
\end{equation}
Here $A$ is the number of nucleons in the reference state, $A^*=A+2$
is the mass number of the target nucleus, and $\op{v}_{ij}$ is 
the nucleon-nucleon interaction. Only two-body interactions are included at present.
In the second quantization, the Hamiltonian can be written as
\begin{equation}
\label{eq:hamilton2}
\op{H}=\sum_{pq} \varepsilon^p_q a^\dagger_p a_q 
+\frac{1}{4}\sum_{pqrs} \langle pq||rs\rangle a^\dagger_p a^\dagger_q a_s a_r \ .
\end{equation}
The term $\langle pq||rs\rangle$ is a shorthand for the matrix elements (integrals) of the two-body part of the Hamiltonian
of Eq.~\eqref{eq:hamilton}, $p,q,r$ and $s$ represent various single-particle states while $\varepsilon^p_q$ stands for the matrix elements of the one-body operator in Eq.~\eqref{eq:hamilton}. 
Finally, second quantized operators like $a^\dagger_q$ and $a_p$ create 
and annihilate a nucleon in the
states $q$ and $p$, respectively.
These operators fulfill the canonical
anti-commutation relations.

\subsection{Single-reference coupled-cluster theory}
In single-reference coupled-cluster theory, the many-body ground-state $|\Psi_0\rangle $ 
is given by the exponential ansatz,  
\begin{equation}
    |\Psi_0\rangle  = \exp{(\op{T})} |\Phi_0\rangle \ .
    \label{eq:cc_exp_anz}
\end{equation}
Here, $|\Phi_0\rangle $ is the reference Slater determinant, where all states below
the Fermi level are occupied and 
$\op{T}$ is the cluster operator that generates correlations. The
operator $\op{T}$ is expanded as a linear combination of
particle-hole excitation operators
\begin{equation}
    \op{T} = \op{T}_1 + \op{T}_2 + \ldots + \op{T}_A \, 
    \label{eq:cc_op_cluster}
\end{equation}
where $\op{T}_n$ is the $n$-particle-$n$-hole($n$p-$n$h) excitation operator
\begin{equation}
  \op{T}_n = \left(\frac{1}{n!}\right)^2 \sum_{a_\nu i_\nu}
  t_{i_1 \ldots i_n}^{a_1 \ldots a_n} a^\dagger_{a_1} \dots a^\dagger_{a_n} 
a_{i_n} \dots a_{i_1} \ .
    \label{eq:cc_op_npnh}
\end{equation}
Throughout this work the indices $i j k\ldots$
denote states below the Fermi level (holes), while the indices
$a b c\ldots$ denote states above the Fermi level (particles). For an
unspecified state, the indices $pqr\ldots$ are used. The amplitudes
$t_{i_1 \ldots i_n}^{a_1 \ldots a_n}$ will be determined by solving
the coupled-cluster equations.  In the singles and doubles approximation 
the cluster operator is truncated as
\begin{equation}
    \op{T} \approx \op{T}_{\rm CCSD} \equiv \op{T}_1 + \op{T}_2 \ ,
    \label{eq:cc_op_ccsd}
\end{equation}
which defines the coupled-cluster approach with singles and doubles excitations, the so-called  
CCSD approximation. The unknown amplitudes result from the solution of the 
non-linear CCSD equations given by
\begin{eqnarray}
\label{eq:cc_eq_amp}
\langle \Phi_i^a|\barh|\Phi_0\rangle &=& 0, \nonumber\\
\langle \Phi_{ij}^{ab}|\barh|\Phi_0\rangle &=& 0 \ .
\end{eqnarray}
The term
\begin{equation}
    \barh = \exp{(-\op{T})}\op{H}_N\exp{(\op{T})} = \left( \op{H}_N \exp{(\op{T})} \right)_C,
    \label{eq:cc_barh}
\end{equation}
is called the similarity-transform of the normal-ordered Hamiltonian.
In this formulation, the state
$|\Phi_{ij\dots}^{ab\dots}\rangle$ is a Slater determinant that differs
from the reference $|\Phi_0\rangle$ by holes
in the orbitals $ij\dots$ and by particles in the orbitals $ab\dots$. The
subscript $C$ indicates that only connected diagrams enter, while the
normal-ordered Hamiltonian is defined as
\begin{equation}
    \op{H}_N = \op{F} + \op{V}_N - E_0,
\end{equation}
The operator $\op{F}$ is the onebody part of the normal-ordered Hamiltonian defined as
\begin{equation}
    \label{eq:onebody}
    \op{F} = \sum_{pq} f_q^p \left\{a^\dagger_p a_q\right\},
\end {equation}
where
\begin{equation}
    \label{eq:f_elements}
    f^p_q = \epsilon_q^p  + \sum_i \bra{pi}\ket{qi}.
\end{equation}
Here $\epsilon_q^p$ and $\bra{pi}\ket{qi}$ are the
matrix elements of the Hamiltonian in Eq.~\eqref{eq:hamilton2}. The sum is over
all single particle indices, $i$, below the Fermi energy.
The operator $\op{V}_N$ is the twobody part of the normal-ordered Hamiltonian,
while $E_0$ denotes the vacuum expectation value with respect to the reference
state.

Once the $t^a_i$ and $t^{ab}_{ij}$ amplitudes have been determined
from Eq.~(\ref{eq:cc_eq_amp}), the correlated ground-state
energy is given by
\begin{equation}
  E_{\rm CC} = \langle\Phi_0|{\barh}|\Phi_0\rangle + E_0 \ . 
    \label{eq:cc_eq_energy}
\end{equation}

The CCSD approximation is a very inexpensive method to obtain the ground state
energy of a nucleus. In most cases however, the accuracy is not
satisfactory~\cite{hagen2008}. The
obvious solution would be to include triples excitations in
Eq.~\eqref{eq:cc_op_ccsd} to define the CCSDT approximation. This leads to
an additional set of non-linear equations
\begin{equation}
    \bra{\Phi_{ijk}^{abc}} \barh \ket{\Phi_0} = 0,
\end{equation}
that has to be solved consistently. Unfortunately, such a calculation is 
computationally prohibitive~\cite{Hagen2007b}. The computational cost of CCSDT scales as
$n_o^3 n_u^5$, where $n_o$ is the number of single-particle states occupied in the reference
determinant and $n_u$ are the number of unoccupied states. For
comparison, the computational cost of the CCSD approximation scales as $n_o^2
n_u^4$.

Instead of solving the coupled-cluster equations~\eqref{eq:cc_eq_amp} including
triples excitations, one calculates a correction to the correlated ground state
energy~\eqref{eq:cc_eq_energy}, using the $\Lambda$-CCSD(T)
approach~\cite{Kucharski1998,Taube2008}. Here, the left-eigenvalue problem using
the CCSD similarity-transformed Hamiltonian is solved, yielding a 
correction to the ground state energy. The left-eigenvalue problem is
given by
\begin{equation}
    \bra{\Phi_0} \op{\Lambda} \barh = E\bra{\Phi_0} \op{\Lambda}
    \label{eq:lt_evp},
\end{equation}
where $\op{\Lambda}$ is a de-excitation operator,
\begin{equation}
    \op{\Lambda} = \op{1} + \op{\Lambda}_1 + \op{\Lambda}_2,
\end{equation}
and
\begin{align}
    \op{\Lambda}_1 &= \sum_{i a} \lambda_a^i a_a a^\dagger_i, \\
    \op{\Lambda}_2 &= \sum_{i j a b} \lambda_{ab}^{ij} a_b a_a a_i^\dagger
    a_j^\dagger.
\end{align}
The unknown amplitudes $\lambda_a^i$ and $\lambda_{ab}^{ij}$ are the components
of the left-eigenvector with the lowest eigenvalue in Eq.~\eqref{eq:lt_evp}.
Once found, the energy correction is given by
\begin{multline}
    \Delta E_3 = \frac{1}{(3!)^2} \sum_{ijkabc} \bra{\Phi_0}
    \op{\Lambda}\left(\op{F}_{hp} + \op{V}_N\right)\ket{\Phi_{ijk}^{abc}} \\
    \times \frac{1}{\epsilon_{ijk}^{abc}} \bra{\Phi_{ijk}^{abc}}
    \left(\op{V}_N \op{T}_2\right)_C \ket{\Phi_0}.
\end{multline}
Here, $\op{F}_{hp}$ is the part of the normal-ordered one-body
Hamiltonian~\eqref{eq:onebody} that
annihilates particles and creates holes. The energy denominator is defined as 
\begin{equation}
    \epsilon_{ijk}^{abc} \equiv f_{ii} + f_{jj} + f_{kk} - f_{aa} - f_{bb} -
    f_{cc},
\end{equation}
where $f_{pp}$ are the diagonal elements of the normal ordered one-body
Hamiltonian $\op{F}$ defined in Eq.~\eqref{eq:f_elements}.

Using this approach, the ground state wave function~\eqref{eq:cc_exp_anz} and the similarity
transformed Hamiltonian~\eqref{eq:cc_barh} are calculated using the CCSD
approximation, while the ground state energy
is given by
\begin{equation}
    E_{\Lambda\textrm{CC}} = E_{\textrm{CC}} + \Delta E_3
\end{equation}
This approximation has proved to give very accurate results for closed (sub-)shell
nuclei~\cite{Hagen2010a}.

\subsection{Equation-of-motion coupled-cluster(EOM-CC) theory}
In nuclear physics, the single reference coupled-cluster method defined by the coupled-cluster
equations\eqref{eq:cc_eq_amp} is normally used to obtain the ground state
energy of a closed (sub-)shell nucleus. While it is possible to apply the CC
method to any reference determinant to obtain  the energy of different states,
the EOM-CC framework is usually employed
for such endeavors.

Equation~\eqref{eq:cc_barh} defines a similarity transformation. This guarantees that
the eigenvalues of $\barh$ are equivalent to the eigenvalues of the intrinsic
Hamiltonian~\eqref{eq:hamilton} and that the eigenvectors are connected by the
transformation defined by Eq.~\eqref{eq:cc_exp_anz}.
However, approximations are introduced by limiting the vector space allowed in
the diagonalization of $\barh$. This is the foundation of the EOM-CC
approach.

To simplify the equations and for effective calculations, the
eigenvalue problem in the EOM-CC approach is modified. A new eigenvalue problem
is defined as the difference between a
target state and the coupled-cluster reference state~\eqref{eq:cc_exp_anz}.
Formally, a general state of the $A$-body nucleus is written
\begin{equation}
    \ket{\Psi_\mu} = \op{\Omega}_\mu \ket{\Psi_0} = \op{\Omega}_\mu
    e^{\op{T}} \ket{\Phi_0}. \label{eq:eom-reference}
\end{equation}
Here $\op{\Omega}_\mu$ is an excitation operator that creates the state
$\ket{\Psi_\mu}$ when applied to the coupled-cluster reference state ~$\ket{\Psi_0}$. The
label $\mu$ identifies the quantum numbers(eg. energy and angular momentum)  of
the target state. The \SE{}s for the target state and the coupled-cluster
reference state
are written
\begin{align}
    \op{H} \op{\Omega}_\mu e^{\op{T}} \ket{\Phi_0} &= E_\mu \op{\Omega}_\mu 
    e^{\op{T}} \ket{\Phi_0} \label{eq:eom_se_full} \\
    \op{H} e^{\op{T}} \ket{\Phi_0} &= E_\textrm{CC} e^{\op{T}} \ket{\Phi_0}.
    \label{eq:eom_se_cc}
\end{align}
Here $E_\mu$ is the energy of the target state and $E_\textrm{CC}$ is the
coupled-cluster reference energy in Eq.~\eqref{eq:cc_eq_energy}.

By multiplying Eq.~\eqref{eq:eom_se_full} with $e^{-\op{T}}$ and
Eq.~\eqref{eq:eom_se_cc} with $\op{\Omega}_\mu e^{-\op{T}}$ from the left and  take the
difference between the two equations, the eigenvalue problem is written as
\begin{equation}
    \left[ \barh, \op{\Omega}_\mu \right] \ket{\Psi_0} = \omega_\mu
        \op{\Omega}_\mu \ket{\Psi_0},
\end{equation}
where $\omega_\mu = E_\mu - E_{\textrm{CC}}$ and we have used that $\smash{\left[
\op{\Omega}_\mu, \op{T}\right] = 0}$.
Finally, none of the unconnected terms in the evaluation of the commutator survive, resulting in
\begin{equation}
    \left( \barh \op{\Omega}_\mu \right)_C \ket{\Phi_0} = \omega_\mu \op{\Omega}_\mu
    \ket{\Phi_0}.
    \label{eq:eom-master}
\end{equation}
This operator equation can be posed as a matrix eigenvalue problem where $\omega_\mu$ are the eigenvalues and the
matrix elements of $\op{\Omega}_\mu$ are the components of the eigenvectors. 
The subscript $C$ implies that only terms where $\barh$ and $\op{\Omega}_\mu$
are connected by at least one contraction survive. In diagrammatic terms, this
means that only connected diagrams appear in the operator product ${\left( \barh
\op{\Omega}_\mu \right)_C}$.

The similarity-transformed Hamiltonian \eqref{eq:cc_barh} is a
non-Hermitian operator and is diagonalized by an Arnoldi algorithm( for details, see, for example, \textcite{golubvanloan}).  This
algorithm relies on the repeated application of the connected matrix vector product defined
by Eq.~\eqref{eq:eom-master}. A left-eigenvalue problem is solved to obtain the
conjugate eigenvectors~\cite{Stanton1993}, but this is beyond the scope of this
article.

To find the explicit expressions for the connected matrix vector product, the
excitation operator must be properly defined.
When used for excited states of an $A$-body nucleus, the excitation
operator in Eq.~\eqref{eq:eom-reference} is parametrized in terms of $n$p-$n$h
operators and written as
\begin{equation}
    \op{\Omega}_\mu = \op{R} = \op{1} + \op{R}_1 + \op{R}_2 + \dots \op{R}_A
    \label{eq:op_eom_es},
\end{equation}
where
\begin{equation}
    \op{R}_n = \frac{1}{(n!)^2} \sum_{\substack{
        i_1, \dots i_n \\
        a_1, \dots a_n}}
        r_{i_1 \ldots i_n}^{a_1 \ldots a_n} a^\dagger_{a_1} \ldots
        a^\dagger_{a_n} a_{i_n}\ldots a_{i_1}.
\end{equation}
The unknown amplitudes $r$(with the sub- and superscripts dropped) are the matrix elements
of $\op{R}$,
\begin{equation}
    r_{i_1 \ldots i_n}^{a_1 \ldots a_n} = 
    \bra{\Phi_{i_1 \ldots i_n}^{a_1 \ldots a_n}} \op{R} \ket{\Phi_0},
    \label{eq:eom-amplitudes}
\end{equation}
and can be grouped into a
vector that solves the eigenvalue problem in Eq.~\eqref{eq:eom-master}. The
explicit equations for the matrix vector product are established by looking at
each individual element using a diagrammatic approach,
\begin{equation}
    \left(\barh \op{R}\right)_{i_1 \ldots i_n}^{a_1 \ldots a_n} \equiv
    \bra{\Phi_{i_1 \ldots i_n}^{a_1 \ldots a_n}} \left(\barh \op{R}\right)_C
    \ket{\Phi_0}.
\end{equation}

Calculations using the full excitation operator~\eqref{eq:op_eom_es}, are not computationally
tractable, so an additional level of approximation is introduced by a
truncation.
When the CCSD approximation is used to obtain
the reference wave function, the excitation operator is truncated at the
$2$p-$2$h level~\cite{Comeau1993} which defines EOM-CCSD.

In the EOM-CC approach, the states of $A\pm k$ nuclei are also treated as excited
states of an $A$-body nucleus. The general wave function for an $A\pm k$ nucleus
is written
\begin{equation}
    \ket{\Psi_\mu^{A\pm k}} = \op{\Omega}_\mu \ket{\Psi_0^{(A)}} = \op{\Omega}_\mu
    e^{\op{T}} \ket{\Phi_0}. \label{eq:eom_2pa_reference}
\end{equation}
The operator $\op{\Omega}_\mu$ and the energies $E_\mu$ of the target state,
also solve the eigenvalue problem in Eq.~\eqref{eq:eom-master}. The energy difference
$\omega_\mu = E_\mu - E_0^*$ is now the excitation energy of the target state in the
nucleus $A\pm k$, with respects to the closed-shell reference nucleus with the
mass shift $A^*=A\pm k$ in the Hamiltonian~\eqref{eq:hamilton}. This mass shift ensures that the correct kinetic energy
of the center of mass is used in computing the $A\pm k$ nuclei.

The operators
\begin{equation}
    \op{\Omega}_\mu = \op{R}^{A\pm 1} = \op{R}^{A\pm 1}_1 + \op{R}^{A\pm 1}_2 + \dots
    \op{R}^{A\pm 1}_A
\end{equation}
where
\begin{align}
    \op{R}^{A+1}_n &= \frac{1}{(n!)(n-1)!} \sum_{\substack{
    i_1, \dots i_{n-1} \\
        a_1, \dots a_n}}
        r_{i_1 \ldots i_{n-1}}^{a_1 \ldots a_n}\nn
        & \quad \times a^\dagger_{a_1} \ldots
        a^\dagger_{a_n} a_{i_{n-1}}\ldots a_{i_1} \\
    \op{R}^{A-1}_n &= \frac{1}{(n!)(n-1)!} \sum_{\substack{
    i_1, \dots i_{n} \\
    a_1, \dots a_{n-1}}}
    r_{i_1 \ldots i_{n}}^{a_1 \ldots a_{n-1}}\nn
        & \quad \times a^\dagger_{a_1} \ldots
        a^\dagger_{a_{n-1}} a_{i_{n}}\ldots a_{i_1}
\end{align}
define the particle attached equation-of-motion coupled-cluster
~\cite{Musial2003b} (PA-EOM-CC)
and the particle removed equation-of-motion coupled-cluster
~\cite{Musial2003a}(PR-EOM-CC)
approaches. These methods have been used successfully in quantum chemistry for
some time (see \textcite{Bartlett2007} for a review), but also have recently been
implemented for use in nuclear structure calculations~\cite{Hagen2010b}.

In \textcite{Jansen2011} 2PA-EOM-CCSD and 2PR-EOM-CCSD were defined for systems
with two particles attached to and removed from a closed (sub-)shell nucleus. For
this problem, the
excitation operators were given by
\begin{equation}
    \op{\Omega}_\mu = \op{R}^{A\pm 2} = \op{R}^{A\pm 2}_2 + \op{R}^{A\pm 2}_3 + \dots
    \op{R}^{A\pm 2}_A \label{eq:op_2pa_full}
\end{equation}
where
\begin{align}
    \op{R}^{A+2}_n &= \frac{1}{(n!)(n-2)!} \sum_{\substack{
    i_1, \dots i_{n-2} \\
        a_1, \dots a_n}}
        r_{i_1 \ldots i_{n-2}}^{a_1 \ldots a_n}\nn
        & \quad \times a^\dagger_{a_1} \ldots
        a^\dagger_{a_n} a_{i_{n-2}}\ldots a_{i_1} \\
    \op{R}^{A-2}_n &= \frac{1}{(n!)(n-2)!} \sum_{\substack{
    i_1, \dots i_{n} \\
    a_1, \dots a_{n-2}}}
    r_{i_1 \ldots i_{n}}^{a_1 \ldots a_{n-2}}\nn
        & \quad \times a^\dagger_{a_1} \ldots
        a^\dagger_{a_{n-2}} a_{i_{n}}\ldots a_{i_1}.
\end{align}
In this article, I will focus on the 2PA-EOM-CCSD method,
where~\eqref{eq:op_2pa_full} is truncated at the $3$p-$1$h level. This approximation is suitable for states with
a dominant $2$p structure. It is already computationally
intensive with up to $10^9$ basis states~(see Sec.~\ref{sec:results}) for the largest
nuclei attempted. A full inclusion of $4$p-$2$h amplitudes therefore is not feasible at this
time.

\subsection{Spherical coupled-cluster theory}
For nuclei with closed (sub-)shell structure, the reference state has good spherical
symmetry and zero total angular momentum. For these systems, the cluster
operator~\eqref{eq:cc_op_cluster} is a scalar under rotation and depends only on
reduced amplitudes.
Thus,
\begin{equation}
    \op{T}_1 = \sum_{i a} t_i^a(J)\left[ a^\dagger_a(J) \otimes
    \tilde{a}_i(J)\right]^{0}
\end{equation}
and
\begin{equation}
    \op{T}_2 = \sum_{i j a b J} t_{ij}^{ab}(J)\left[\left[
        a^\dagger_a(j_a) \otimes a^\dagger_b(j_b)\right]^J \otimes
        \left[\tilde{a}_j(j_j) \otimes \tilde{a}_i(j_i)\right]^J \right]^0,
\end{equation}
where the amplitudes $t(J)$(sub- and superscripts dropped) are a short form of the
reduced matrix elements of the cluster operator~\eqref{eq:cc_op_cluster} (see Appendix~\ref{app:reduced} for details).
Moreover, $J$ is a label specifying the total angular momentum of a many-body state
and standard tensor notation has been used to specify the tensor couplings. The
single particle operator $\tilde{a}_i$ is the time reversal of the $a^\dagger_i$
operator that creates a particle in the orbital labeled $i$.

As the similarity-transformed Hamiltonian~\eqref{eq:cc_barh} is a product of three
scalar operators~(remember that the exponential of an operator is defined in
terms of its Taylor expansion), it is also a scalar under
rotation. This allows a formulation of the coupled-cluster equations that is
completely devoid of magnetic quantum numbers, thus reducing the size of
the single-particle space and the number of
coupled non-linear equations to solve in Eq.~\eqref{eq:cc_eq_amp}.
For further details, see \textcite{Hagen2010a}.

Within the same formalism, the connected operator product
in Eq.~\eqref{eq:eom-master} is established. This will greatly reduce the computational cost of
calculating the product but also allow a major reduction in both the
single-particle basis and the number of allowed configurations in the many-body
basis.

Given a target state with total angular momentum $J$ (in units of $\hbar c$),
the excitation operator, $\Omega_\mu$\eqref{eq:eom-reference}, is a spherical
tensor operator by
definition~(see, for example, \textcite{Bohr1969}). It has a rank of $J$, with $2J+1$
components labeled by the magnetic quantum number $M \in [-J, \ldots, J]$. It
is written as
\begin{equation}
    \Omega_\mu = \op{R}_\mu^{A\pm k}(J,M) \label{eq:eom_op_reduced},
\end{equation}
where $A$ is the number of particles in the reference state, $A\pm k$ is the
number of particles in the target state, while $\mu$ identifies a specific set
of quantum numbers.
Identifying the excitation operator as a spherical tensor operator, invokes an extensive
machinery of angular momentum algebra with important theorems. Of special
importance is the \WE(see for example \textcite{edmunds1960}), which states that the matrix
elements of a spherical tensor operator can be factorized into two parts. The
first is a geometric part identified by a \CG, while the second is a
reduced matrix element that does not depend on the magnetic quantum
numbers.

To develop the spherical form of EOM-CC, I will use the following notation
for the matrix elements of a general operator
\begin{equation}
    \bra{ab} \op{O} \ket{ij} \equiv 
    \bra{\Phi^{ab}_{ij}} \op{O} \ket{\Phi_0},
\end{equation}
where the single-particle states labeled $a$ and $b$ are occupied in the
outgoing state, while the single-particle states labeled $i$ and $j$ are occupied in the
incoming state. All single-particle states shared between the incoming and
outgoing many-body states are dropped from the notation.

In this form, a component of the spherical basis is written as
\begin{equation}
    \ket{\alpha; J_\alpha M_\alpha},
\end{equation}
where $\alpha$ denotes a particular many-body state, while $J_\alpha$($M_\alpha$) is the
total angular momentum(projection) of this state. 
Using the spherical notation, the matrix elements of the excitation operator are written
\begin{equation}
        r_\beta^\alpha(J_\alpha, J_\beta) =
        \bra{\alpha; J_\alpha M_\alpha} \op{R}^J_M \ket{\beta; J_\beta
            M_\beta},
\end{equation}
where we have dropped the cumbersome sub- and superscripts on the excitation
operator in favor of standard tensor notation.
The matrix elements of the matrix vector product in
Eq.~\eqref{eq:eom-master} are written
\begin{multline}
    \bra{\alpha; J_\alpha M_\alpha} \left(\barh \op{R}^J_M \right)_C
    \ket{\beta; J_\beta M_\beta} = \\
    \omega  \bra{\alpha; J_\alpha M_\alpha} \op{R}^J_M \ket{\beta; J_\beta
    M_\beta}.
\end{multline}

Now the \WE{} allows a factorization of the matrix elements into two factors
\begin{multline}
    C^{J J_\beta J_\alpha}_{M M_\beta M_\alpha} 
    \bra{\alpha; J_\alpha }| \left(\barh \op{R}^J \right)_C |
    \ket{\beta; J_\beta } = \\
    \omega C^{J J_\beta J_\alpha}_{M M_\beta M_\alpha}
    \bra{\alpha; J_\alpha }| \op{R}^J |\ket{\beta; J_\beta }.
\end{multline}
Here $C^{J J_\beta J_\alpha}_{M M_\beta M_\alpha}$ is a \CG{} and the double
bars denote reduced matrix elements and do not depend on any of the projection
quantum numbers. This equation is simplified by dividing by the \CG{}. 
This means that for each set of $\alpha$, $\beta$, $J_\alpha$, and
$J_\beta$, where $J$, $J_\alpha$, and $J_\beta$ satisfy the triangular
condition, there are $(2J+1)\times (2J_\alpha+1) \times (2J_\beta+1)$
identical equations for a given $J$. Only one is needed to solve the eigenvalue
problem, which reduces the dimension of the problem significantly.
In the final eigenvalue problem the
unknown components of the eigenvectors are the reduced matrix elements of the
excitation operator
\begin{multline}
    \bra{\alpha; J_\alpha}| \left(\barh \op{R}^J \right)_C |
        \ket{\beta; J_\beta } =
    \omega \bra{\alpha; J_\alpha }| \op{R}^J |
    \ket{\beta; J_\beta }. \label{eq:eom_sphmaster}
\end{multline}
The eigenvalue problem in Eq.~\eqref{eq:eom_sphmaster} is the spherical
formulation of the general EOM-CC diagonalization problem. For a given
excitation operator, both the connected operator product and the reduced
amplitudes must be defined explicitly.

\section{\label{sec:2pa}Spherical 2PA-EOM-CCSD}
In this work I derive the spherical formulation of the 2PA-EOM-CCSD~\cite{Jansen2011} method, where the excitation operator
in Eq.~\eqref{eq:op_2pa_full} has been truncated at the $3$p-$1$h
level. It is defined as
\begin{equation}
    \op{R} = \frac{1}{2} \sum_{ab} r^{ab} a^\dagger_a a^\dagger_b +
    \frac{1}{6} \sum_{abci} r^{abc}_i a^\dagger_a a^\dagger_b a^\dagger_c a_i,
    \label{eq:eom_operator}
\end{equation}
where the cumbersome sub- and superscripts in the operator have been dropped. 

Let us begin by introducing the notation used throughout this section.
The unknown amplitudes $r$ are the matrix elements of $\op{R}$ and defined
by
\begin{align}
    r^{ab} &= \bra{\Phi^{ab}} \op{R} \ket{\Phi_0} \equiv \bra{ab}\op{R}\ket{0}\nn
    r^{abc}_i &= \bra{\Phi^{abc}_i} \op{R} \ket{\Phi_0} \equiv \bra{abc} \op{R} \ket{i},
\end{align}
while a shorthand form of the components of the matrix-vector product is
introduced
\begin{align}
    \left(\barh \op{R}\right)^{ab} &= \bra{\Phi^{ab}}\left(\barh
    \op{R}\right)_C\ket{\Phi_0}\label{eq:eom_mvp_2pa} \\
    \left(\barh \op{R}\right)^{abc}_i &= \bra{\Phi^{abc}_i}\left(\barh
    \op{R}\right)_C\ket{\Phi_0}.\label{eq:eom_mvp_3p1h}
\end{align}

In this notation, the eigenvalue problem in Eq.~\eqref{eq:eom-master} is written
\begin{align}
    \left(\barh \op{R}\right)^{ab} &= \omega  r^{ab} \nn
    \left(\barh \op{R}\right)^{abc}_i &= \omega r^{abc}_i.
    \label{eq:eom_2pa_master}
\end{align}

In the spherical formulation, the excitation operator is a spherical tensor
operator of rank $J$ and projection $M$,
\begin{multline}
    \op{R}^J_M = 
    \frac{1}{2}
    \sum_{ab} r^{ab}(J) 
    \left[ a^\dagger_a(j_a) \otimes  a^\dagger_b(j_b) \right]^J_M \\
    \quad
    + 
    \frac{1}{6}
        \sum_{\substack{
        abci \\
        J_{ab} J_{abc}}}
        r^{abc}_i\left(J, J_{abc}, J_{ab}\right) \\
        \times\left[\left[
        \left[a^\dagger_a(j_a) \otimes  a^\dagger_b(j_b)\right]^{J_{ab}}
        \otimes a^\dagger_c(j_c) \right]^{J_{abc}} \otimes
        \tilde{a}_i(j_i)\right]^J_M. \label{eq:eom_operator_sph}
\end{multline}
Here the $a^\dagger_a(j_a)$ and $\tilde{a}_i(j_i)$ are spherical tensor
operators of rank $j_a$ and $j_i$ respectively, where the latter is the
time-reversed operator of $a^\dagger_i(j_i)$. 
Standard tensor
notation has been used to define the spherical tensor couplings.
The reduced amplitudes are now the reduced matrix elements of the spherical
excitation operator~\eqref{eq:eom_operator_sph}. They are defined as
\begin{equation}
    r^{ab}(J) = \bra{ab; j_a j_b; J}|\op{R}^J|\ket{0},
    \label{eq:eom_amplitude_2pa}
\end{equation}
where $j_a$ and $j_b$ are coupled to $J$ in left to right order. Moreover,
\begin{equation}
    r^{abc}_i\left(J, J_{abc}, J_{ab}\right) =
    \bra{abc; j_a j_b; J_{ab} j_c; J_{abc}}|\op{R}^J|\ket{i; j_i},
\end{equation}
where $j_a$ and $j_b$ has been coupled to $J_{ab}$,
while $J_{ab}$ and $j_c$ has been coupled to $J_{abc}$, also in left to right order. 
The shorthand form of
the reduced matrix elements of the connected operator product is defined
analogously by
\begin{equation}
    \left(\barh \op{R}^J\right)^{ab}(J) = \bra{ab; j_a j_b; J}|\left(\barh
    \op{R}^J\right)_C|\ket{0} \label{eq:eom_mvp_2pa_sph}
\end{equation}
and
\begin{multline}
    \left(\barh \op{R}^J\right)^{abc}_i\left(J, J_{abc}, J_{ab}\right) = \\
    \bra{abc; j_a j_b; J_{ab} j_c; J_{abc}}|\left(\barh
    \op{R}^J\right)_C|\ket{i; j_i}.
\end{multline}
The transformations that connect the reduced matrix elements of $\op{R}^J$ with
the uncoupled matrix elements are given in
Eqs.~\eqref{eq:reduced_2p}-\eqref{eq:reduced_3p1h_rev}.

The final form of the spherical eigenvalue problem~\eqref{eq:eom_sphmaster} is written 
\begin{align}
    \left(\barh \op{R}^J\right)^{ab}(J) &= \omega  r^{ab}(J) \nn
    \left(\barh \op{R}^J\right)^{abc}_i(J, J_{abc}, J_{ab}) &= \omega
    r^{abc}_i(J, J_{abc}, J_{ab}),
    \label{eq:eom_2pa_sphmaster}
\end{align}
where the amplitudes are the reduced matrix elements defined above.

\begin{table*}[ht]
    \begin{ruledtabular}
\begin{tabular}{lll}
    Diagram & Uncoupled expression & Coupled expression \\ 
    \hline
    \\
        \includegraphics[scale=0.4]{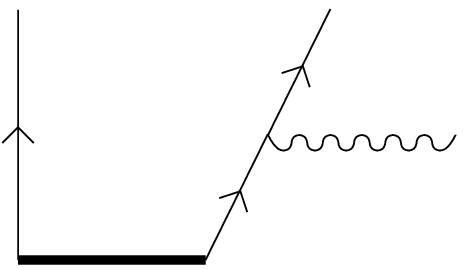} & 
        $\op{P}(ab)
        {\barh}_e^b r^{ae}$ &
        $\op{P}(ab)\barh_e^b(j_b)r^{ae}(J)$ \\
        \includegraphics[scale=0.4]{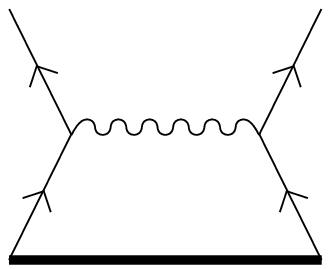} & 
        $\frac{1}{2}
        \barh_{ef}^{ab} r^{ef}$ &
        $\frac{1}{2} \barh_{ef}^{ab} (J) r^{ef}(J)$ \\
        \includegraphics[scale=0.4]{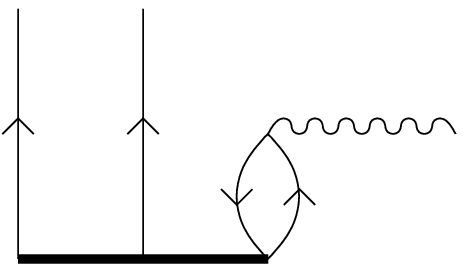} & 
        ${\barh}_{e}^m r_m^{abe}$ &
        $\sum_{J_{abe}} \barh_e^m(j_e) r^{abe}_m(J_{ab}, J_{abe}, J)
        \frac{\hat{J}_{abe}^2}{\hat{J}^2}$ \\
        \includegraphics[scale=0.4]{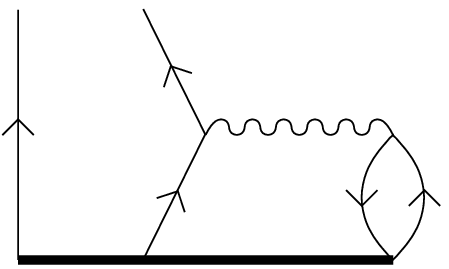} & 
        $\frac{1}{2}
        \op{P}(ab) {\barh}_{ef}^{bm} r_m^{aef}$ &
        $\frac{1}{2} \op{P}(ab) \sum_{J_{efb}, J_{ef}} 
        (-1)^{1 + j_b + j_m -J_{ef} - J} 
        \frac{\hat{J}^2_{efb} \hat{J}_{ef}}{\hat{J}}
        \begin{Bmatrix}
            j_b & j_a & J \\
            j_m & J_{efb} & J_{ef}
        \end{Bmatrix}$ \\
        &&
        $\times
        \barh_{ef}^{am}(J_{ef})
        r^{efb}_m(J_{ef}, J_{efb}, J)$ \\
        \includegraphics[scale=0.4]{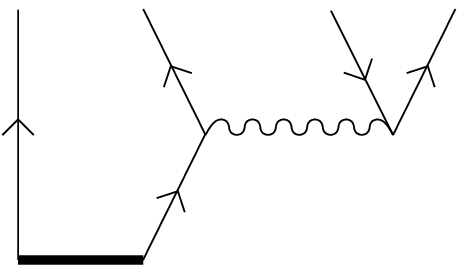} &
        $\op{P}(a,bc)
        {\barh}_{ei}^{bc} r^{ae}$ &
        $\op{P}(ab,c) (-1)^{1 + j_c + j_i +J_{ab} - J} \hat{J}_{ab} \hat{J}
        \begin{Bmatrix}
            j_c & j_e & J \\
            j_i & J_{abc} & J_{ab}
        \end{Bmatrix}
        \barh_{ab}^{ei}(J_{ab}
        r^{ec}(J)$ \\
        \includegraphics[scale=0.4]{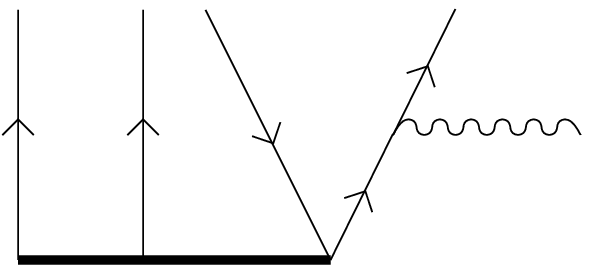} &
        $\op{P}(ab,c)
        {\barh}_e^c r_i^{abe}$ &
        $\op{P}(ab,c) \barh_e^c(j_c) r^{abe}_i(J_{ab}, J_{abe}, J)$ \\
        \includegraphics[scale=0.4]{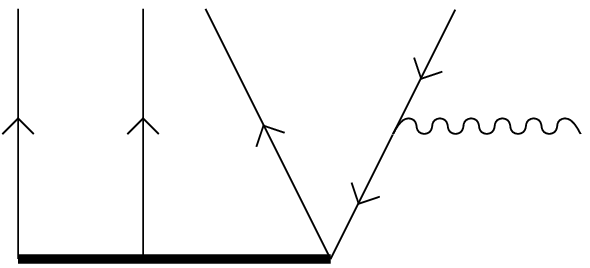} & 
        $-{\barh}_i^m  r_m^{abc}$ &
        $- \barh_m^i(j_i)
        r^{abc}_m(J_{ab}, J_{abc}, J)$ \\
        \includegraphics[scale=0.4]{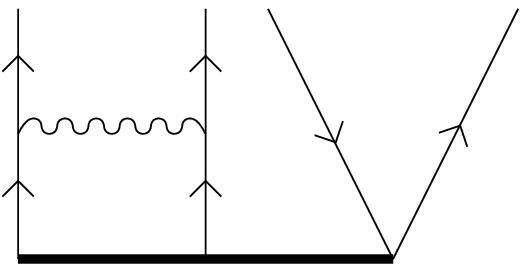} & 
        $\frac{1}{2} \op{P}(ab,c) {\barh}_{ef}^{ab} r_i^{efc}$ &
        $\frac{1}{2}
        \op{P}(ab,c)
        \barh_{ef}^{ab}(J_{ab})
        r^{efc}_i(J_{ab}, J_{abc}, J)$ \\
        \includegraphics[scale=0.4]{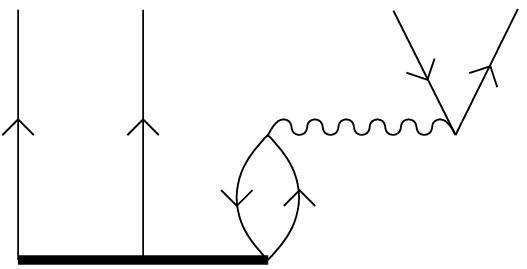} &
    $\op{P}(ab,c){\barh}_{ei}^{mc} r_m^{abe}$ &
    $\op{P}(ab,c)
    \sum_{J_{abe}, J_{mc}}
    (-1)^{1 + j_e + j_m + J_{abe} + J_{abc} + J_{mc}}
    \hat{J}_{abe}^2 \hat{J}_{mc}^2
    \begin{Bmatrix}
        J_{ab} & j_e & J_{abe} \\
        j_c & J_{mc} & j_m \\
        J_{abc} & j_i & J
    \end{Bmatrix}$ \\
    &&
    $\times
    \barh_{ei}^{mc}(J_{mc})
    r^{abe}_m(J_{ab}, J_{abe}, J)$ \\
        \includegraphics[scale=0.4]{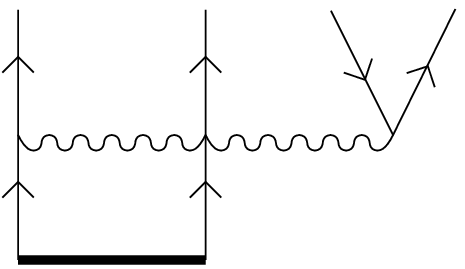} &
        $\frac{1}{2}
        {\barh}_{efi}^{abc} r^{ef}$ &
    \multirow{2}{*}{
    $(-1)^{1+j_c+j_m -J}
    \hat{J} \hat{J}_{ab}
    \begin{Bmatrix}
        j_c & j_m & J \\
        j_i & J_{abc} & J_{ab}
    \end{Bmatrix}
    \chi_m^c(J)
    t_{im}^{ab}(J_{ab})$} \\
        \includegraphics[scale=0.4]{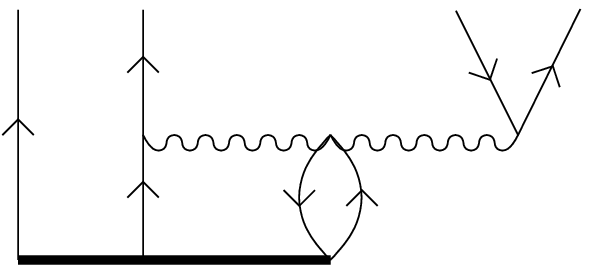} &
        $\frac{1}{2} \hat{P}(a,bc) {\barh}^{bmc}_{efi} r_m^{aef}$
\end{tabular}
\end{ruledtabular}
\caption{\label{tab:2pa_all}All diagrams for the 2PA-EOM-CCSD method with both
ordinary and reduced amplitudes and matrix elements. The reduced amplitudes and
matrix elements are defined in Appendix~\ref{app:reduced}, while $\chi_i^a(J)$ is
defined in Appendix~\ref{app:threebody}. Note that the two last diagrams are
combined into one expression in the spherical formulation and that repeated
indices are summed over.}
\end{table*}

Table~\ref{tab:2pa_all} presents the main result of this section. The first
column lists all possible diagrams that contribute to the matrix-vector product
in Eqs.~\eqref{eq:eom_2pa_master} and \eqref{eq:eom_2pa_sphmaster}. The
remaining two columns contain the closed form
expressions for these diagrams in the uncoupled and in the spherical
representation respectively. All matrix elements and amplitudes are defined in
Appendix~\ref{app:reduced}, while the permutation operators $\op{P}(a,b)$ and
$\op{P}(ab,c)$ are defined in Appendix~\ref{app:permutation}. Note that in the
spherical representation the permutation operators also change the coupling
order.

The last two diagrams contain the three-body parts of the similarity-transformed
Hamiltonian~\eqref{eq:cc_barh} and have been combined in the spherical
representation. The details of how the intermediate operator $\op{\chi}$ is
defined, are contained in Appendix~\ref{app:threebody}.

Let us briefly go through the derivation of a single spherical diagram
expression. The first diagram in Table~\ref{tab:2pa_all} will serve as a good example.
This diagram contributes to the $2$p matrix elements defined in
Eq.~\eqref{eq:eom_mvp_2pa} in the uncoupled representation and to the reduced
matrix elements defined in Eq.~\eqref{eq:eom_mvp_2pa_sph} in the spherical
representation.

The first step is to use the transformation in Eq.~\eqref{eq:reduced_2p} to
write the reduced matrix elements~\eqref{eq:eom_mvp_2pa_sph} in terms of the
uncoupled matrix elements~\eqref{eq:eom_mvp_2pa}. This gives us 
\begin{equation}
    \left(\barh \op{R}^J\right)^{ab}(J) = \frac{1}{\hat{J}^2} \sum_{M m_a m_b}
    C^{j_a j_b J}_{m_a m_b M} \left(\barh \op{R}\right)^{ab},
    \label{eq:eom_mvp_transf}
\end{equation}
where $\hat{J} \equiv \sqrt{2J+1}$. The diagram
contributions to the uncoupled matrix elements are given by
\begin{equation}
    \left(\barh \op{R}\right)^{ab} \gets \op{P}(ab) {\barh}_e^b r^{ae},
    \label{eq:d1_contrib}
\end{equation}
where the arrow indicates that it
is only one of several contributions to this matrix element. Here, ${\barh}_e^b$
is a matrix element of the one-body part of the similarity-transformed
Hamiltonian~\eqref{eq:cc_barh} and both ${\barh}_e^b$ and $r^{ae}$ are in the
uncoupled representation.
        
Second, Eq.~\eqref{eq:d1_contrib} is inserted into Eq.~\eqref{eq:eom_mvp_transf}
to get
\begin{equation}
    \left(\barh \op{R}\right)^{ab}(J) \gets 
        \frac{1}{\hat{J}^2} 
        \sum_{M m_a m_b}
        C^{j_a j_b J}_{m_a m_b M}
        {\barh}_e^b r^{ae}.
\end{equation}
Note that for the moment, we are ignoring the permutation operator $\op{P}(ab)$
that is a part of the diagram. 

Third, the reverse transformations in Eqs.~\eqref{eq:reduced_barh1_rev}
and \eqref{eq:reduced_2p_rev} are used to transform the uncoupled matrix elements of ${\barh}_e^b$
and $r^{ae}$ to the corresponding reduced matrix elements. This
gives
\begin{multline}
        \left(\barh \op{R}\right)^{ab}(J) \gets 
        \frac{1}{\hat{J}^2} 
        \sum_{M}
        {\barh}_e^b(j_b) r^{ae}(J)
        \\ 
        \times
        \sum_{ m_a m_b}
        \delta_{j_b, j_e} \delta_{m_b, m_e}
        C^{j_a j_b J}_{m_am_b M}
        C^{j_a j_e J}_{m_am_e M},
\end{multline}
where $\delta$ is the Kronecker $\delta$ and comes from the
application of the \WE{} to the matrix element of $\barh$. The \CG s are
orthonormal so
\begin{equation}
        \sum_{ m_a m_b}
        \delta_{j_b, j_e} \delta_{m_b, m_e}
        C^{j_a j_b J}_{m_am_b M}
        C^{j_a j_e J}_{m_am_e M} = 1.
\end{equation}

The remaining expression simplifies to 
\begin{equation}
    \left(\barh \op{R}\right)^{ab}(J) \gets
        {\barh}_e^b(j_b) r^{ae}(J).
\end{equation}
Note that $\sum_M 1 = 2J+1$ and that repeated indices are summed over.

Initially, we left out the permutation operator $\op{P}(a,b)$ that is needed to
generate antisymmetric amplitudes. In the uncoupled representation this operator
is defined as
\begin{equation}
    \op{P}(ab) = \op{1} - \op{P}_{a,b},
\end{equation}
where $\op{1}$ is the identity operator and $\op{P}_{a,b}$ changes the
order of the two indices $a$ and $b$, but leaves the coupling order unchanged.
Let us apply this operator to $\left(\barh \op{R}\right)^{ab}(J)$. The result is
\begin{multline}
    \op{P}(a,b) \left(\barh \op{R}\right)^{ab}(J) = \\
    \left(\barh \op{R}\right)^{ab}(J) - 
    \bra{ba; j_a j_b; J}|\left(\barh \op{R}\right)_C|\ket{0},
\end{multline}
where the last matrix element has the wrong coupling order compared to the
reduced amplitudes defined in Eq.~\eqref{eq:eom_mvp_2pa_sph} where
\begin{equation}
    \left(\barh \op{R}\right)^{ba}(J) = 
        \bra{ba; j_b j_a; J}|\left(\barh \op{R}\right)_C|\ket{0}.
\end{equation}
To change the coupling order, one of the symmetry properties of the \CG s is
exploited to write
\begin{multline}
    \bra{ba; j_a j_b; J}|\left(\barh \op{R}\right)_C|\ket{0} = \\
    (-1)^{j_a + j_b - J} \bra{ba; j_b j_a; J}|\left(\barh
    \op{R}\right)_C|\ket{0} \\
    = (-1)^{j_a + j_b - J} \left(\barh \op{R}\right)^{ba}(J).
\end{multline}

To simplify the notation, the permutation operator in the spherical
representation is defined to also change the coupling order. This results in the following
definition
\begin{equation}
    \op{P}(ab) = \op{1} - (-1)^{j_a + j_b - J} \op{P}_{a,b}.
    \label{eq:op_perm_2p}
\end{equation}

The total contribution from the first diagram in Table~\ref{tab:2pa_all} in the
spherical representation is given by
\begin{equation}
    \left(\barh \op{R}\right)^{ab}(J) \gets
        \op{P}(ab) {\barh}_e^b(j_b) r^{ae}(J),
\end{equation}
where $\op{P}(ab)$ is defined by Eq.~\eqref{eq:op_perm_2p}.

The three-body permutation operators $\op{P}(ab,c)$ are defined in the same
manner, but they must change the coupling order of three angular momenta. The
details have been left to Appendix~\ref{app:permutation}.

\section{\label{sec:results}Results}
\subsection{Model space and interaction}
All calculations in this section have been done in a spherical Hartree-Fock
basis, based on harmonic oscillator single-particle wave functions. These are
identified with the set of quantum numbers $\{nlj\}$ for both protons and
neutrons, where $n$ represents the number of nodes, $l$
represents the orbital momentum, and finally $j$ is the
total angular momentum of the single-particle wave function.

The size of the model space is identified by the variable
\begin{equation}
    \nmax = \textrm{max}(N), \label{eq:res_nmax}
\end{equation}
where $N= 2n + l$, so the number of harmonic oscillator shells is $\nmax +
1$. All single-particle states with
\begin{equation}
    2n+l \leq \nmax
\end{equation}
are included and no additional restrictions are made on the allowed
configurations.
Thus, $\nmax$ completely determines the
computational size and complexity of the calculations.

\begin{table}[htp]
    \centering
    \begin{ruledtabular}
    \begin{tabular}{cccc}
        $N_{\textrm{max}}$ & Size & Elements & Memory \\
        \hline
        10 & 132 & \numprint{145623788} & $1.1$ Gb\\
        12 & 182 & \numprint{587531302} & $4.4$ Gb\\
        14 & 240 & \numprint{1963734704} & $14.6$ Gb\\
        16 & 306 & \numprint{5687352954} & $42.4$ Gb\\
        18 & 380 & \numprint{14715230212} & $109$ Gb \\ 
        20 & 458 & \numprint{33622665364} & $250$ Gb \\
    \end{tabular}
    \end{ruledtabular}
    \caption{\label{tab:size_interaction}The first column contains the size of the single-particle basis
    employed for different model spaces labeled by $\nmax$(See text for details). Column two and
    three list the number of matrix elements for the different model spaces and
    the memory footprint of the interaction in our implementation. All numbers
    are based on the coupled representation, also known as $jj$-scheme. }
\end{table}

Table~\ref{tab:size_interaction} lists the size of the single-particle
space for different values of $\nmax$ in the spherical representation.
In addition, it includes the total number of matrix elements of 
the interaction in Eq.~\eqref{eq:hamilton}, as well as the memory footprint in
the implementation.
Given the memory requirements, it is
clear that a distributed storage scheme is needed.

In addition to the interaction elements, the Arnoldi
vectors in the diagonalization procedure also has to be stored. Typically $150$ iterations are
performed, where one vector has to be stored for each iteration.
\begin{table*}
    \begin{ruledtabular}
    \npthousandsep{ }
\begin{tabular}{ccccccc}
    State & $\nmax = 10$ & $\nmax = 12$ & $\nmax = 14$ & $\nmax = 16$ & $\nmax = 18$ & $\nmax = 20$ \\
    \hline
    ${}^{6}$He$(0^+)$ & \numprint{516048}  & \numprint{1323972}  & \numprint{2981930}  & \numprint{6088376}  & \numprint{11513088}  & \numprint{20176104}  \\
    ${}^{6}$He$(1^-)$ & \numprint{1507930} & \numprint{3894028}  & \numprint{8808688}  & \numprint{18040354} & \numprint{34190482}  & \numprint{60011982}  \\
    ${}^{6}$He$(2^+)$ & \numprint{2391692} & \numprint{6251128}  & \numprint{14255896} & \numprint{29364090} & \numprint{55885624}  & \numprint{98356664} \\
    \\
    ${}^{6}$Li$(0^+)$ & \numprint{775992}  & \numprint{1989508}  & \numprint{4478936}  & \numprint{9142216}  & \numprint{17284308}  & \numprint{30285212}  \\
    ${}^{6}$Li$(1^+)$ & \numprint{2268746} & \numprint{5853534}  & \numprint{13234004} & \numprint{27093632} & \numprint{51335514}  & \numprint{90080136}  \\
    ${}^{6}$Li$(2^+)$ & \numprint{3595384} & \numprint{9391650}  & \numprint{21409878} & \numprint{44088456} & \numprint{83893672}  & \numprint{147629532} \\
    ${}^{6}$Li$(3^+)$ & \numprint{4676372} & \numprint{12438258} & \numprint{28699916} & \numprint{59604726} & \numprint{114125048} & \numprint{201657602} \\
    \\
    ${}^{18}$O$(0^+)$ & \numprint{1908474}  & \numprint{5022710}  & \numprint{11485808}  & \numprint{23680034}  & \numprint{45071990}  & \numprint{79331610}  \\
    ${}^{18}$O$(1^-)$ & \numprint{5594899}  & \numprint{14802528}  & \numprint{33974801}  & \numprint{70231288}  & \numprint{133940727}  & \numprint{236049974}  \\
    ${}^{18}$O$(2^+)$ & \numprint{8891923}  & \numprint{23794936}  & \numprint{55036119}  & \numprint{114391274}  & \numprint{219038683}  & \numprint{387077788}  \\
    ${}^{18}$O$(2^-)$ & \numprint{8897760}  & \numprint{23803219}  & \numprint{55047530}  & \numprint{114406595}  & \numprint{219058796}  & \numprint{387083193}  \\
    ${}^{18}$O$(3^+)$ & \numprint{11613562}  & \numprint{31596862}  & \numprint{73906056}  & \numprint{154840950}  & \numprint{298237942}  & \numprint{529098382}  \\
    ${}^{18}$O$(3^-)$ & \numprint{11621868}  & \numprint{31608838}  & \numprint{73922708}  & \numprint{154863424}  & \numprint{298267524}  & \numprint{529107862}  \\
    ${}^{18}$O$(4^+)$ & \numprint{13629562}  & \numprint{37905214}  & \numprint{89982332}  & \numprint{190504054}  & \numprint{369757342}  & \numprint{659327780}  \\
    \\
    ${}^{18}$F$(0^+)$ & \numprint{2868568}  & \numprint{7545420}  & \numprint{17248686}  & \numprint{35552756}  & \numprint{67658660}  & \numprint{119071548}  \\
    ${}^{18}$F$(1^+)$ & \numprint{8403602}  & \numprint{22228738}  & \numprint{51009366}  & \numprint{105427688}  & \numprint{201040066}  & \numprint{354285892}  \\
    ${}^{18}$F$(2^+)$ & \numprint{13362878}  & \numprint{35742012}  & \numprint{82642970}  & \numprint{171734254}  & \numprint{328788766}  & \numprint{580957010}  \\
    ${}^{18}$F$(3^+)$ & \numprint{17451568}  & \numprint{47458334}  & \numprint{110973350}  & \numprint{232452890}  & \numprint{447659068}  & \numprint{794095862}  \\
    ${}^{18}$F$(4^+)$ & \numprint{20479376}  & \numprint{56930198}  & \numprint{135106850}  & \numprint{285982274}  & \numprint{554996372}  & \numprint{989530134}  \\
    ${}^{18}$F$(5^+)$ & \numprint{22363324}  & \numprint{63896228}  & \numprint{154444460}  & \numprint{331158558}  & \numprint{648765300}  & \numprint{1163943530}  \\
\end{tabular}

    \end{ruledtabular}
    \caption{\label{tab:vectorsize_2pa}Size of the many-body space in the
    diagonalization procedure in the Arnoldi algorithm for all states calculated
    in this work. All numbers 
        are based on the angular-momentum coupled representation($jj$-scheme).}
\end{table*}
Table~\ref{tab:vectorsize_2pa} lists the size of a single vector for
selected target states in various model spaces. As an example, for a double precision
calculation, where each element requires $8$ bytes of storage, the Arnoldi
diagonalization would require $\approx 76$~GB of memory for the
$J^\pi=3^+$ state of ${}^6$Li with $\nmax = 16$. Thus the Arnoldi
procedure quickly becomes the largest memory consumer in this method.
In general, there is a large computational cost from increasing the total angular momentum of the target
state, comparable to increasing the size of the model space.

The interaction used in this work is derived from chiral perturbation theory at
next-to-next-to-next-to-leading order(\nlo) using the interaction matrix elements of
\textcite{Entem2003}. The matrix elements of this interaction employs a cutoff
$\Lambda=500$~MeV and all partial waves up to relative angular
momentum $J_{rel}= 6$ are included. The relevant three- and
four-body interactions defined by the chiral expansion at this order are not
included.

For the treatment of center-of-mass contamination, a softer interaction where
the short-range parts are removed via the similarity renormalization group
transformation~(SRG)~\cite{Bogner2007}, is used. A cutoff
$\lambda=2.0\textrm{fm}^{-1}$ is sufficient for this purpose.

\subsection{\label{sec:com}Treatment of center of mass}
Recently, \textcite{Hagen2009b,Hagen2010a} demonstrated a procedure to show that the coupled-cluster wave function
separates into an intrinsic part and a Gaussian for the center-of-mass
coordinate. This is important, because the model spaces employed in
coupled-cluster calculations are not  complete $N \hbar \omega$ spaces, where the
basis sets consist of all $A$-body Slater determinants not exceeding $N \hbar \omega$
in excitation energy. In practical calculations, where the model spaces are not
complete, the separation therefore is not \emph{a priori} guaranteed. As a
result, the intrinsic Hamiltonian, where all reference to the center-of-mass has
been removed, is usually employed.

In the EOM-CC approach, one makes further approximations by truncating the many-body
basis before a diagonalization is performed. It therefore is not clear that
the final wave functions separate in the same way as the coupled-cluster
reference state. In the following, I will investigate the center-of-mass
properties of 2PA-EOM-CC wave functions. As an example, I will highlight selected
solutions for $A=6$ nuclei. First, we review the procedure from
\textcite{Hagen2009b,Hagen2010a} and introduce the notation.

First, it is assumed that the wave function is the $n$'th eigenvalue of the
center-of-mass Hamiltonian 
\begin{equation}
    \op{H}^{(n)}_{cm}(\tilde{\omega}) = \op{T}_{cm} + \frac{1}{2} m A
    \tilde{\omega}^2
    \op{R}_{cm}^2 -\left(\frac{3}{2} + n\right) \hbar \tilde{\omega} \label{eq:hcom},
\end{equation}
with a frequency $\hbar \tilde{\omega}$ that, in general, differs from that of the harmonic
oscillator basis employed in the calculation. The expectation value of this
operator should vanish given the correct value of $n$. For all physical
solutions, the expectation should vanish for $n=0$, provided the
solutions are converged. This
assumption is rooted in the observation that for most coupled-cluster wave
functions, the
expectation value $E^{(0)}_{cm}(\omega) = \langle \op{H}^{(0)}_{cm}(\omega)
\rangle$ is, in general, not zero for different
values of $\hbar \omega$ but close to zero for a specific value. Further, there
seems to be very little correlation between $E^{(0)}_{cm}(\omega)$ and the energy
of the coupled-cluster solution. Although the coupled-cluster solution is
completely independent of $\hbar \omega$, $E^{(0)}_{cm}(\omega)$ is not.

Second, one demands that the expectation value $E^{(n)}_{cm}(\tilde{\omega}) =
\langle \op{H}^{(n)}_{cm}(\tilde{\omega})\rangle$
vanishes for a given value of $\hbar \tilde{\omega}$, independent of $\hbar
\omega$. Under this requirement, the numerical value of $\hbar \tilde{\omega}$
is given by
\begin{multline}
    \hbar \tilde{\omega} = \hbar \omega + \frac{2}{2n+1} E^{(n)}_{cm}(\omega) \pm \\
        \sqrt{\frac{4}{(2n+1)^2} \left(E^{(n)}_{cm}(\omega)\right)^2 + \frac{4}{2n+1}\hbar
        \omega E^{(n)}_{cm}(\omega)}, \label{eq:hwtilde}
\end{multline}
that only depends on the frequency of
the harmonic oscillator basis employed the calculation.

Finally, one calculates the expectation value
$E^{(n)}_{cm}(\tilde{\omega})$, where $\hbar \tilde{\omega}$ now depends
on $\hbar \omega$. If $\hbar \tilde{\omega}$ is constant with respect to $\hbar
\omega$ and $E^{(n)}_{cm}(\tilde{\omega}) \approx 0$ for a range of $\hbar
\omega$ values, the original assumption is verified.

In the following, I will present results for the $J=0^+$ ground state of ${}^{6}$He
and the first excited $J=3^+$ state of ${}^{6}$Li. In addition I include a low
lying $J^\pi = 1^-$ state that shows up in the numerical spectrum of ${}^{6}$He. This
state has not been documented experimentally and is a prime candidate for a
spurious center-of-mass excitation.
All calculations were performed in a model space defined by $\nmax = 16$, which
was sufficient for converged energies for all states, using an SRG transformed interaction
with a momentum cutoff $\lambda = 2.0$~fm${}^{-1}$. 

\begin{figure}[htp]
    \centering
    \includegraphics[width=0.45\textwidth]{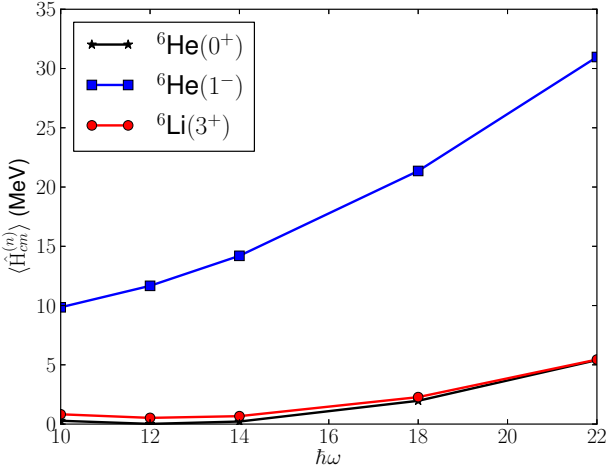}
    \caption{\label{fig:hcm_energy}(Color online) Expectation value of the center-of-mass
    Hamiltonian~\eqref{eq:hcom} at the frequency $\hbar \tilde{\omega} = \hbar
    \omega$.  Three different states are shown -- the $J^\pi=0^+$ ground state of
    ${}^{6}$He, the $J^\pi = 1^-$ excited state of ${}^{6}$He and the
    $J^\pi=3^+$ excited state of ${}^{6}$Li. All states are assumed to be the
    lowest eigenstates of the center-of-mass Hamiltonian in Eq.~\eqref{eq:hcom}
    with $n=0$.} 
\end{figure}

Figure~\ref{fig:hcm_energy} shows the expectation value of the center-of-mass
Hamiltonian~\eqref{eq:hcom} at the frequency $\hbar \tilde{\omega} = \hbar
\omega$ for the three states in question. It is assumed that all states are
degenerate with the ground state with $n=0$. The $J^\pi = 0^+$ state of ${}^{6}$He
and the $J^\pi=3^+$ state of ${}^{6}$Li shows the expected behavior as observed in
\textcite{Hagen2009b,Hagen2010a}. The expectation value vanishes for $\hbar \omega \approx 12$~MeV, but not in general. The
expectation value with respect to the $J^\pi=1^-$ state however, does not vanish
for any frequency. It is clearly wrong to assume that it is the ground state
of the center-of-mass Hamiltonian~\eqref{eq:hcom}.

Instead, let us assume that it is the first excited state of the center-of-mass
Hamiltonian~\eqref{eq:hcom} with $n=1$ . This would make it a $p$ state
with negative parity which gives a $J^\pi = 1^-$ state when coupled to a
$J^\pi=0^+$ intrinsic state. If this is the case, it will be a spurious center-of-mass
excitation where the intrinsic wave function is degenerate with the intrinsic
ground state. 
\begin{figure}[htp]
    \centering
    \includegraphics[width=0.45\textwidth]{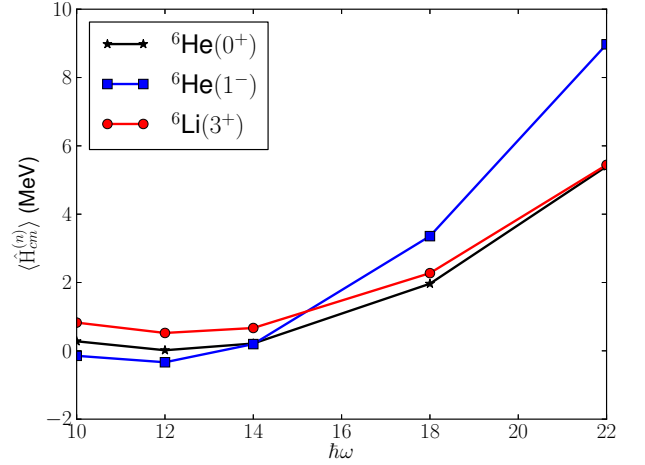}
    \caption{\label{fig:hcm_energy_ex}(Color online) Expectation value of the center-of-mass
    Hamiltonian~\eqref{eq:hcom} at the frequency $\hbar \tilde{\omega} = \hbar
    \omega$.  Three different states are shown -- the $J^\pi=0^+$ ground state of
    ${}^{6}$He, the $J^\pi = 1^-$ excited state in ${}^{6}$He and the
    $J^\pi=3^+$ excited state in ${}^{6}$Li.
    The positive parity states are assumed to be the lowest eigenstates of the
    center-of-mass Hamiltonian~\eqref{eq:hcom} with $n=0$, while the negative
    parity state is assumed to be the first excited state of the center-of-mass
    Hamiltonian~\eqref{eq:hcom} with $n=1$.}
\end{figure}

Figure~\ref{fig:hcm_energy_ex} shows the same information as
Fig.~\ref{fig:hcm_energy}, only now the $J^\pi=1^-$ state in ${}^{6}$He is
assumed to be the first excited state of the center-of-mass
Hamiltonian~\eqref{eq:hcom} with $n=1$. The expectation value now vanishes for all three
states at $\hbar \omega \approx 12$~MeV.

\begin{figure}[htp]
    \centering
    \includegraphics[width=0.45\textwidth]{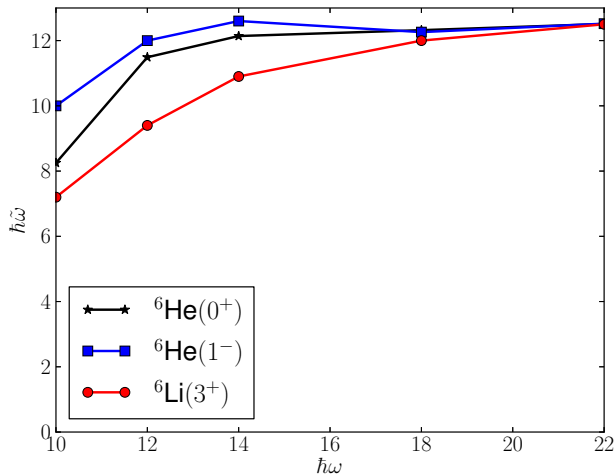}
    \caption{\label{fig:hwt}(Color online) Center-of-mass frequency as calculated by the
    description in \textcite{Hagen2009b,Hagen2010a} as a function of the oscillator
    parameter $\hbar \omega$. Three different states are shown -- the $J^\pi=0+$
    ground state of ${}^{6}$He, the $J^\pi = 1^-$ excited state in ${}^{6}$He
    and the $J^\pi=3^+$ excited state in ${}^{6}$Li.
    The positive parity states are assumed to be the lowest eigenstates of the
    center-of-mass Hamiltonian~\eqref{eq:hcom} with $n=0$, while the negative
    parity state is assumed to be the first excited state of the center-of-mass
    Hamiltonian with $n=1$.}
\end{figure}

Under these assumptions, the appropriate $\hbar \tilde{\omega}$ is calculated
using  Eq.~\eqref{eq:hwtilde}. As seen in Fig.~\ref{fig:hwt},
where $\hbar \tilde{\omega}$ is plotted as a function of $\hbar \omega$, the
frequency of the center-of-mass Hamiltonian is approximately independent of the
frequency of the underlying harmonic oscillator basis for all three states.

\begin{figure}[htp]
    \centering
    \includegraphics[width=0.45\textwidth]{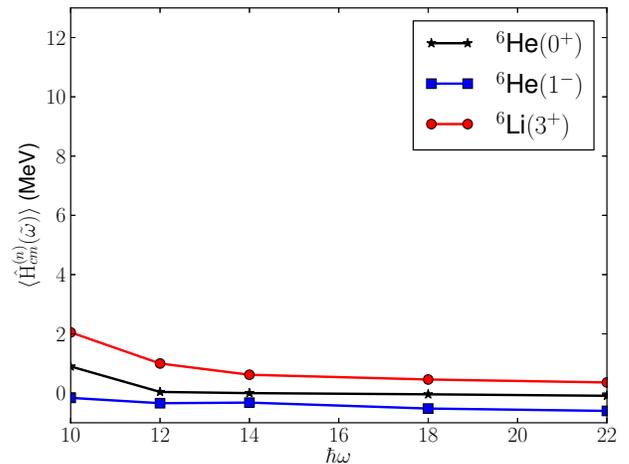}
    \caption{\label{fig:hwt_energy}(Color online) The expectation value of the center-of-mass
    Hamiltonian~\eqref{eq:hcom} is calculated at the center-of-mass frequency
    $\hbar \tilde{\omega}$
    for the $J^\pi=0+$ ground state of ${}^{6}$He, the $J^\pi = 1^-$ excited state in
    ${}^{6}$He and the $J^\pi=3^+$ excited state in ${}^{6}$Li. 
    The positive parity states are assumed to be the lowest eigenstates of the
    center-of-mass Hamiltonian~\eqref{eq:hcom} with $n=0$, while the negative
    parity state is assumed to be the first excited state of the center-of-mass
    Hamiltonian with $n=1$.}
\end{figure}

Finally, the  expectation values are calculated using
Eq.~\eqref{eq:hcom}. The results are shown in Fig.~\ref{fig:hwt_energy}.

From these results, we can draw a couple of conclusions.
First, since the expectation values are approximately zero, this shows that our
assumptions were valid. All states are approximate
eigenstates of the center-of-mass Hamiltonian~\eqref{eq:hcom}. This means that the
total wave function separates
into an intrinsic part and a center-of-mass part for all three states.
Second, the wave functions for the ground state of ${}^{6}$He and the
first excited $J^\pi=3^+$ state of ${}^{6}$Li, factorizes into intrinsic states and the
ground state of the center-of-mass Hamiltonian. Last, the $J^\pi=1^-$ state in ${}^{6}$He factorizes into
the intrinsic ground state and the first excited state of the center-of-mass
Hamiltonian. It is identified as a spurious center-of-mass excitation and should be removed from
the spectrum.

Ideally, one should go through the entire procedure outlined above to make sure
that the calculated state is not a spurious center-of-mass excitation.
In practice, it is only necessary to verify that the center-of-mass energy
$E^{(0)}_{cm}(\hbar \omega)$ vanishes for some value of $\hbar \omega$. 

There are several reasons why the results in this section are only
approximate. First, the method used to calculate expectation values is not
exact. Second, the single-particle space employed in the calculations is cut off
at some maximum energy. Although it is verified that the total energy is
converged with respect to this cutoff, properties of the wave function might
require higher cutoffs. Third, the results obtained by the coupled-cluster machinery are
truncated both in the coupled-cluster expansion and in the operator used to
define the diagonalization space. Finally, the interaction used in these
calculations has been evolved using SRG transformations. The three- and
many-body forces induced by this transformation have not been included in these
calculations. If one assumes that the first two items yield small deviations
from zero, then it might be possible to use this to evaluate how good the
coupled-cluster truncations are and say something about the current level of
approximation. However, one will need to incorporate $4$p-$2$h corrections to
analyze this further.

The purpose of this section has been to show that it is possible to identify and exclude spurious center-of-mass
excitations for both ground and excited states calculated with EOM-CC theory. The wave function factorizes, to a very good
approximation, into an intrinsic part and a harmonic oscillator eigenfunction for the
center-of-mass coordinate. To deternine why this is the case will require additional research
and is beyond the scope of this article.

\subsection{Applications to ${}^{6}$Li and ${}^{6}$He}
For any given reference nucleus, there are only three nuclei accessible to the
2PA-EOM-CC method. Using ${}^{4}$He as the reference, one can add two protons to calculate
properties of ${}^6$Be, two neutrons for ${}^6$He and finally a proton and a neutron to
calculate properties of ${}^6$Li. Of these, only ${}^6$Li and ${}^6$He are stable with respect
to nucleon emission and will be the focus of this section. 
The structures of  ${}^6$Li and ${}^6$He differ markedly. This is important,
because the quality of the current level of approximation will inevitably
depend on the structure of
the nucleus under investigation.

${}^6$Li is well bound and has four bound states below the nucleon emission threshold at
$4.433$~MeV~\cite{Tilley2002}. The ground state has spin parity assignment $J^\pi
= 1^+$, while the first excited states have $J^\pi=3^+$, $2^+$ and $0^+$. The
$J^\pi=0^+$ ground state in ${}^6$He has a two neutron halo structure, bound by only
$800$~keV~\cite{Tilley2002} compared to ${}^{4}$He. There are no bound excited states,
only a narrow resonance at $1.710$~MeV~\cite{Tilley2002}, and recently also
resonances at $2.6$ and $5.3$~MeV~\cite{Mougeot2012}have been documented.

First, let us look at convergence with respect to the size of the model space.
Coupled-cluster theory is based on a finite basis expansion, where $\nmax$ effectively
determines
the numerical cutoff.  The cutoff is increased until the corrections are so small
that the uncertainties in the method dominate the error budget. Typically the
corrections are down to a tenth of a percentage of the total binding energy.
Extrapolations to infinite model spaces~\cite{Furnstahl2012,More2013} have not
been performed here but will be included in future work.

\begin{figure}[htp]
    \centering
        \includegraphics[width=0.45\textwidth]{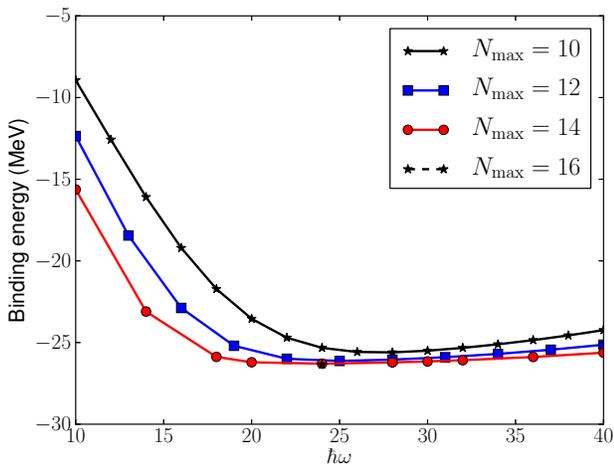}
        \caption{\label{fig:li6_1+0_vbare_j6}(Color online) Ground state energy of ${}^6$Li as
        a function of the oscillator parameter $\hbar \omega$ and the size of the
        model space $\nmax$~(see text for details).}
\end{figure}

Figure~\ref{fig:li6_1+0_vbare_j6} shows the calculated total binding energy of
${}^6$Li as a function of the oscillator frequency $\hbar \omega$.  Different lines
correspond to different model spaces. At $\nmax=16$, there is a shallow minimum around $\hbar
\omega = 24$~MeV and, in a $10$-MeV range including this minimum, the binding energy
varies by approximately $100$~keV. This is less than half a percentage of the total energy.
At low frequencies, the energy deviates substantially from the minimum,
due to the lack of resolution in the single-particle space.

Note that the gain in binding energy when going from $\nmax=14$ to $\nmax=16$ is
also very small, about $40$~keV. 
The binding energy of ${}^6$Li is converged with respect to the size
of the model space ($\nmax$) and the energy at $\hbar \omega = 24$~MeV will be
tabulated. 

The picture is largely identical for the binding energy of ${}^6$He, only the minimum
in energy occurs at $\hbar \omega = 20$~MeV. Here the difference in energy
between the two largest model space is about $140$~keV.

\begin{figure}[htp]
    \centering
        \includegraphics[width=0.45\textwidth]{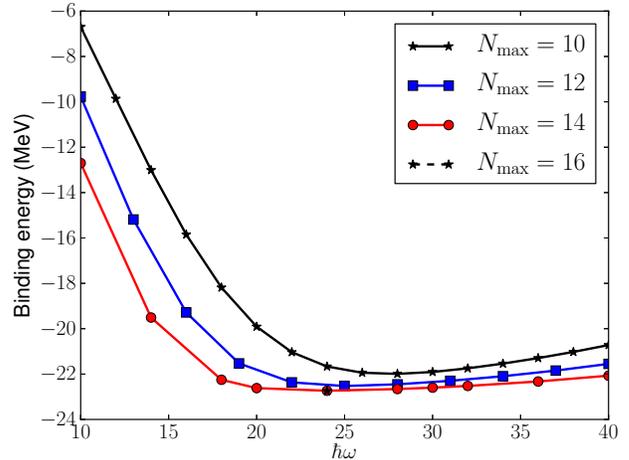}
        \caption{\label{fig:li6_3+1_vbare_j6}(Color online) The total energy of the first
        excited $J^\pi = 3^+$ state of ${}^6$Li as a function of the oscillator
        parameter $\hbar \omega$ and the size of the model space $\nmax$~(see text
        for details).}
\end{figure}
Figure~\ref{fig:li6_3+1_vbare_j6} shows that the excited states follow the same pattern of convergence as
the ground states. Here the total energy of
the $J^\pi=3^+$ state in ${}^6$Li is plotted as a typical example. As before, the energy
is plotted  as a function of $\hbar \omega$ and different lines correspond to
different values of $\nmax$.
In this section, the excitation energy will be defined as
\begin{equation}
    E_{\rm x} (\hbar \omega) = E_{J^\pi}(\hbar \omega) - E_{\rm gs}(\hbar
    \omega),
\end{equation}
where $E_{J^\pi}(\hbar \omega)$ is the total energy of the excited state with
spin-parity assignment $J^\pi$, calculated at the oscillator frequency $\hbar
\omega$. Moreover, $E_{\rm gs}(\hbar \omega)$ is the ground-state energy calculated
at the same frequency.

\begin{figure}[htp]
    \centering
        \includegraphics[width=0.45\textwidth]{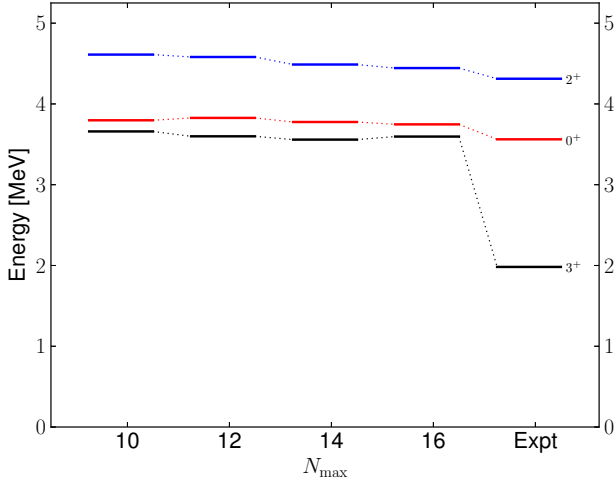}
        \caption{\label{fig:leveltrack_li6}(Color online) Excitation energy for selected states
        of ${}^6$Li, as a function of the size of the model space defined by $\nmax$. The
        rightmost column shows the experimental values from \textcite{Tilley2002}.}
\end{figure}

\begin{figure}[htp]
    \centering
        \includegraphics[width=0.45\textwidth]{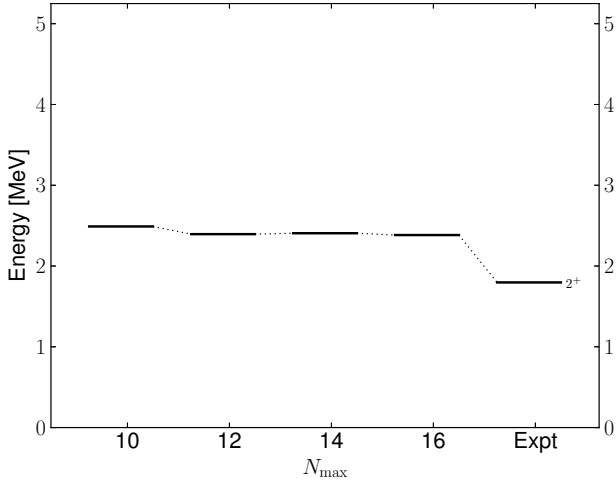}
        \caption{\label{fig:leveltrack_he6}Excitation energy for the first excited $J^\pi = 2^+$ state of 
        of ${}^6$He, as a function of the size of the model space defined by $\nmax$. The
        rightmost column shows the experimental value from \textcite{Tilley2002}.}
\end{figure}
In Fig.~\ref{fig:leveltrack_li6}, the convergence pattern of the excitation
energies for selected states in the spectrum of ${}^6$Li is shown. The horizontal axis denotes
the size of the model space, where the values in the rightmost column are the experimental
values~\cite{Tilley2002}. All excitation energies have been calculated at $\hbar
\omega = 24$~MeV, which correspond to the minimum of the ground-state energy.
There is very little model space dependence at $\nmax = 16$ and none of the
states shown are classified as spurious center-of-mass excitations
according to the prescription in Sec.~\ref{sec:com}.
A second $J^\pi = 1^+$ state was found higher in the spectrum, but this state
was found to be a spurious state and was therefore excluded.

In Fig.~\ref{fig:leveltrack_he6}, an equivalent plot for the first $J^\pi = 2^+$
excited state in ${}^6$He is shown. This result is also converged with respect to the size of the
model space. No significant center-of-mass contamination was found in either this state
or the ground state. As already discussed a low-lying $J^\pi=1^-$ state was
also found, but was identified as a spurious center-of-mass excitation.
Note that all excitation energies for
${}^6$He were calculated at $\hbar \omega = 20$~MeV.

Let us also look at some properties of the wave function. Although it is
not an observable, other expectation values might be more sensitive to changes
in the wave function than the energy.

First, the partial norms are defined by
\begin{align}
    n(2p0h) &= \frac{1}{2}\sum_{ab} (2J+1) \left(r^{ab}(J)\right)^2 \label{eq:norm_2p0h} \\
    n(3p1h) &= \frac{1}{6}\sum_{\substack{
        abci \\
        J_{ab} J_{abc}}}
        \left( 2J_{abc} + 1\right) \left(r_i^{abc}(J_{ab}, J_{abc}, J)\right)^2
        \label{eq:n3p1h},
\end{align}
where $n(2p0h) + n(3p1h) = 1$. The amplitudes $r^{ab}(J)$ and $r_i^{abc}(J_{ab},
J_{abc}, J)$ are the spherical amplitudes defined in
Eqs.~\eqref{eq:reduced_2p} and \eqref{eq:reduced_3p1h}, respectively, while $J_x$ are angular momentum labels. Note that
the angular momentum factors are included so the partial norms are
consistent between the coupled and uncoupled schemes. These norms quantify the
part of the wave function in $2$p-$0$h and $3$p-$1$h configurations,
respectively. Note also that they differ in how they are defined from those used in
\textcite{Hagen2012b}, where the $\frac{1}{2}$ and $\frac{1}{6}$ prefactors were
not used. This gave a larger $3$p-$1$h norm than those in this work, due to a
significant overcounting of the $3$p-$1$h amplitudes.

Second, the total weights are defined by
\begin{align}
    w^{ab}_{pw} &= \frac{2J + 1}{2} \sum_{a,b}\left[\left(r_{pw}^{ab}(J)\right)^2 +
    \left(r_{pw}^{ba}(J)\right)^2 \right] \label{eq:weights_2p0h},
\end{align}
where the label $pw$ identifies the partial wave content of the weight.
The sum is over all configurations with this partial wave content, because the
weights of individual configurations are not stable.
In addition, spin-orbit partners are not distinguished. 

\begingroup
\squeezetable
\begin{table}[htp]
    \begin{ruledtabular}
        \begin{tabular}{ccll}
            State & $n(2p0h)$ & Dominant configuration(s) & Weight(s) \\
            \hline
            ${}^6$Li($1^+$) & $0.91$ & $(p)^2$, $(d)^2$ &  $0.83$, $0.02$ \\
            ${}^6$Li($3^+$) & $0.90$  & $(p)^2$,$(pf)$  & $0.81$, $0.04$ \\
            ${}^6$Li($0^+$) & $0.88$ & $(p)^2$ &  $0.87$ \\
            ${}^6$Li($2^+$) & $0.91$  & $(p)^2$, $(pf)$, $(d)^2$ & $0.77$,
            $0.07$, $0.02$ \\
            ${}^6$He($0^+$) & $0.88$ & $(p)^2$ & $0.87$ \\
            ${}^6$He($2^+$) & $0.90$ & $(p)^2$, $(pf)$ & $0.87$, $0.02$ \\
            ${}^6$He($1^-$) & $0.84$ & $(ps)$, $(pd)$ & $0.49$, $0.34$
        \end{tabular}
    \end{ruledtabular}
    \caption{\label{tab:res_a6_weights} This table shows the $2$p-$0$h partial
    norms~\eqref{eq:norm_2p0h}, as well as the dominant configurations for
    calculated states in both ${}^6$Li and ${}^6$He. The weights are calculated
    according to Eq.~\eqref{eq:weights_2p0h} where all nodes for a given partial
    wave contribute to  the sum and spin-orbit partners are not distinguished. }
\end{table}
\endgroup
In Table~\ref{tab:res_a6_weights} partial norms and dominant weights of selected states in ${}^6$Li and ${}^6$He
are listed.
A few comments are in order. 
First, all physical states are consistent with the shell-model picture, where the
dominant contributions to the wave function come from two valence nucleons in
the $p$ shell. Only the $J^\pi=1^-$ state in ${}^{6}$He contains contributions from the
$sd$ shell, but this is natural as no pure $p$ shell configuration will give a
negative-parity state. It is also a spurious center-of-mass excitation and is
excluded from the spectrum. 
Second, the $2$p-$0$h norm for all physical states are around $0.9$. Only the
spurious state has a significantly lower norm at $0.84$. The remaining $0.1$ in
the $3$p-$1$h norms are needed to relax the reference wave function as it changes
 due to the presence of the extra nucleons.
Finally, the wave function of the ground state of ${}^6$He and the first excited
$J^\pi=0^+$ state in ${}^6$Li are very similar. This is not surprising, since
they can be viewed as two parts of a degenerate isospin triplet.

\begin{table}
    \begin{ruledtabular}
    \begin{tabular}{ccccc}
    \multirow{2}{*}{${}^6$Li} & \multirow{2}{*}{Expt.} & \nlo & NCSM\cite{Navratil2004} \\
    & & ($\Lambda=500\mathrm{MeV}$) \\
    \hline
    $\mathrm{E_{gs}}(1^+)$ & $-31.993$ & $-28.44(5)$ & $-28.5(5)$\\
    \hline
    $\mathrm{E_{gs}}(1^+)$ & $0.0$ & $0.0$ & $0.0$\\
    $\mathrm{E_x}(3_1^+)$ & $+2.186$ & $+3.60(4)$ & $+2.91(3)$\\
    $\mathrm{E_x}(0_1^+)$ & $+3.562$ & $+3.76(3)$ & $+3.30(10)$\\
    $\mathrm{E_x}(2_1^+)$ & $+4.312$ & $+4.44(4)$ & $+4.10(15)$\\
    \hline
    \hline
    \multirow{2}{*}{${}^6$He} \\
    \\
    \hline
    $\mathrm{E_{gs}}(0^+)$ & $-29.270$ & $-25.51(14)$ & $-26.2(5)$ \\
    \hline
    $\mathrm{E_{gs}}(0^+)$ & $0.0$ & $0.0$ & $0.0$\\
    $\mathrm{E_x}(2^+)$ & $1.797$ & $2.35(4)$ \\
\end{tabular}

    \end{ruledtabular}
    \caption{\label{tab:result_A6}Binding energies and excitation energies for
    selected states of ${}^6$Li and ${}^6$He with estimated numerical
    uncertainties. The uncertainties in our results are the differences between
    the values for the two largest model spaces. The data is compared to NCSM
    results~\cite{Navratil2004} in the rightmost column, where the parenthesis 
    list extrapolation errors and experimental  data in the leftmost column. All
    experimental data are from \textcite{Tilley2002}.}
\end{table}
Table~\ref{tab:result_A6} shows results with estimated numerical
uncertainties for the ground and selected excited
states of both ${}^6$He and ${}^6$Li. 
For comparison, both experimental values and results from a 
no-core shell-model (NCSM) calculation~\cite{Navratil2004} are tabulated where data are available.
Note that the results from the NCSM calculation are based on the same
interaction as the results from this work, but the interaction is renormalized
using the procedure defined in \textcite{LeeSuzuki} before the diagonalization was
performed. In addition, the final results were extrapolated to an infinite
model space.

Let us discuss the uncertainties indicated by the parenthesis in the table. For
the results from this work, listed in the second column, the numbers in parenthesis give the difference
in energy between the two largest model spaces. The results from
\textcite{Navratil2004} give the extrapolation errors in the last column,
while the experimental energies~\cite{Tilley2002} in the first column are listed without uncertainties.

\begin{figure}[htp]
    \centering
        \includegraphics[width=0.45\textwidth]{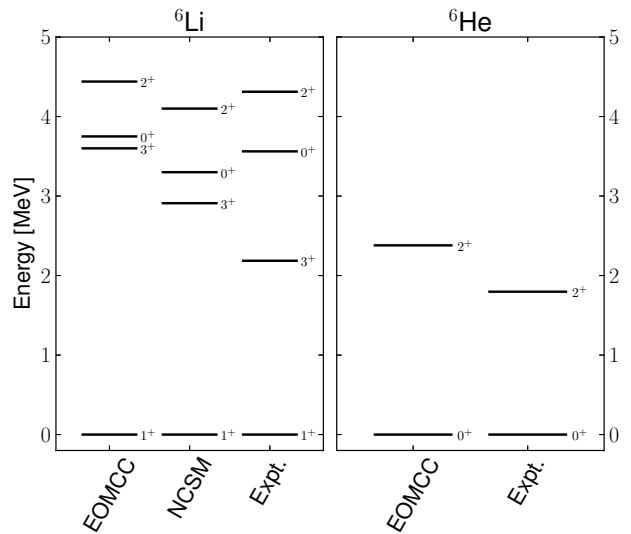}
        \caption{\label{fig:levels_A6}Excitation levels of selected states in 
        ${}^6$Li and ${}^6$He, calculated using 2PA-EOM-CC(this work) and
        NCSM~\cite{Navratil2004}, compared to experimental~\cite{Tilley2002}
        values.}
\end{figure}
Figure~\ref{fig:levels_A6} shows a graphical representation of the data in
Table~\ref{tab:result_A6}.
Compared to the results from the NCSM calculation, our results are quite promising.
First, the ground-state
energy of ${}^6$Li is well within the uncertainties of the ``exact'' result, while
the ground-state energy of ${}^6$He is just outside. The difference between the two
nuclei can be explained by the extended spatial distribution of ${}^6$He. Additional
correlations are necessary to account for this structure.
Although the $\alpha$ core in ${}^6$He is expected to stay largely unchanged when
adding two neutrons, the distribution of these extra neutrons are biased in one
direction. This results in a skewed center of mass compared to the center of mass 
of the $\alpha$ core alone. Additional correlations are necessary to absorb
the resulting oscillations of the $\alpha$ core with respect to the combined
center of mass. The spatial distribution of ${}^6$Li is tighter, so this effect is not that
prominent.
Second, the ordering of excited states in ${}^6$Li is
reproduced.
Finally, the excitation energies are consistently overestimated.
For the first $J^\pi=0^+$ and $2^+$ states the differences between the two
calculations are small enough to be ascribed to differences in the interaction
used. But the difference for the $J^\pi=3^+$ state, however, is too large for
such a simple explanation.
Neither
the partial norms nor the total weights listed in
Table~\ref{tab:res_a6_weights} provide any hint of explanation for this
discrepancy. About $90$~\%  of the wave function is in $2$p-$0$h
configurations, which is comparable to the ground state. The wave function is dominated
by configurations where both nucleons are in $p$ orbitals, which is
consistent with the shell-model picture. Furthermore, the level of convergence
for this state is no different from the other excited states. 
As noted in Sec.~\ref{sec:com}, with the SRG evolved interaction, the
$J^\pi=3^+$ state had a slight center-of-mass contribution, which was not
present in the other states. 
This was illustrated in
Fig.~\ref{fig:hwt_energy}, but a similar calculation using the bare interaction
was too computationally intensive to extract any meaningful information.
I include it because it might indicate that additional
correlations are needed in the calculation, either in the reference or the EOM
operator. This matter needs to be investigated further, but currently the
implementation will not allow model spaces large enough for a converged
description of the center-of-mass admixture in the final wave function.

Using the in-medium similarity renormalization group~(IM-SRG),
\textcite{Tsukiyama2012} performed a similar study with a softer interaction. Here, the $J^\pi=3^+$ state in
${}^6$Li is reproduced on the same level of accuracy as for the other bound states.

Let us also look at some of the differences between the results in this work and the experimental
data. First, all excitation energies are overestimated compared to
data. Again, the $J^\pi=3^+$ states is exceptional, but this has been discussed
in detail by \textcite{Navratil2004}. The matter was resolved by the inclusion
of three-nucleon forces~\cite{Navratil2007}, which also brought the binding
energy very close to data.

There is also an $\approx 500$-keV difference for the $J^\pi=2^+$ resonance
in ${}^6$He, but here the effects of three-body forces might
be less important. This state was also investigated using a chiral interaction with a different cutoff
of $600$~MeV. With this interaction, the excitation energy of this state was
unchanged. That was not the case for the excited states in ${}^6$Li, where especially the
$J^\pi=3^+$ state turned out to be very cutoff dependent.
Since the $J^\pi=2^+$ state in ${}^6$He is a resonance, the continuum is expected to
have a larger impact.
The current single-particle basis cannot
handle the description of both bound, resonance and continuum states
that are necessary in this case.
These effects have not been included in this calculation, as the
focus has been on properties of the method rather than the interaction.
The method has been extended to include a Gamow basis as in
\textcite{Michel2004} and \textcite{Hagen2007c}.  It has already been applied to
${}^{26}$F~\cite{Lepailleur2013}, but a comprehensive discussion is beyond the
scope of this article.

Summing up this section, I would like to point out that for well-bound states, with simple
structure, the current approximation will yield total energies comparable to
exact diagonalization. 
The calculations can be done in sufficiently large
model spaces for the results to be converged for six nucleons, but for certain
states, the effects of $4$p-$2$h configurations need to be investigated.
To compare to experimental data, however, both three-nucleon
forces and continuum degrees of freedom are necessary.

\subsection{Applications to ${}^{18}$O and ${}^{18}$F}
When ${}^{16}$O is used as a reference state, the three isobars ${}^{18}$O, ${}^{18}$F, and ${}^{18}$Ne are
reachable by the 2PA-EOM-CCSD method. All have well-bound ground states and a rich spectra of
bound excited states below their respective nucleon emission thresholds. The
spectrum of ${}^{18}$F is especially rich, as the exclusion principle does not affect the
placement of nucleons in the $sd$ shell. The proton-neutron interaction is
responsible for the compressed spectrum in the fluorine isotope, while strong
pairing effects in ${}^{18}$O
result in a lower ground-state energy.
In both ${}^{18}$O and ${}^{18}$Ne the spectra are opened up and the
first excited states are higher in energy. 
Our current focus is on convergence and the viability of this method. Thus,
${}^{18}$Ne is not explicitly
discussed, as results are similar to those of ${}^{18}$O.

\begin{figure}[htp]
    \centering
        \includegraphics[width=0.45\textwidth]{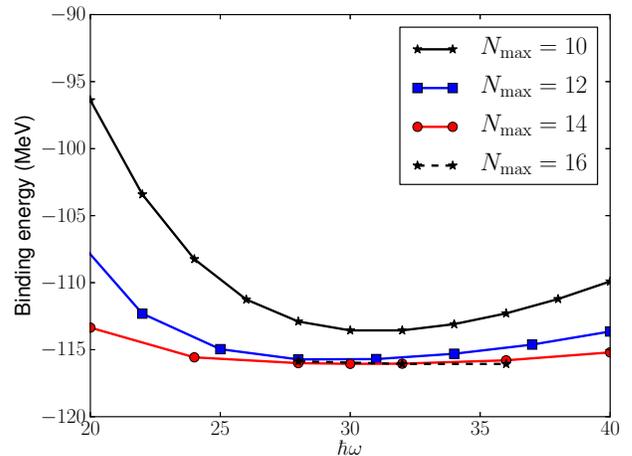}
        \caption{\label{fig:o18_0+0_vbare_j6}(Color online) The ground-state energy of ${}^{18}$O as a
        function of the oscillator parameter, $\hbar \omega$. Different lines
        correspond to different model spaces, parametrized by the variable
        $\nmax$~\eqref{eq:res_nmax}.}
\end{figure}
Let us first look at the convergence of the binding energy of ${}^{18}$O.
Figure~\ref{fig:o18_0+0_vbare_j6} shows the ground-state energy of ${}^{18}$O as a
function of the oscillator parameter $\hbar \omega$. The different
lines correspond to different model spaces, parametrized by the variable
$\nmax$~\eqref{eq:res_nmax}. A shallow minimum develops around $\hbar \omega =
32$~MeV, where the energy is converged with respect to the size of the
model space.
The difference in energy is about $20$~keV when the size of
the model space is increased from $\nmax=14$ to $\nmax=16$.
For a wide range of values around the minimum, the ground state
energy shows very little dependence on the $\hbar \omega$ parameter. Thus, the
result is converged with respect to the size of the model space.

A similar result is
obtained for the ground-state energy of ${}^{18}$F, where the difference in
energy is around $160$~keV between the two largest model spaces. This is almost
an order of magnitude larger than for the ground state of ${}^{18}$O but is still well
within $1\%$ of the total energy.

\begin{figure}[htp]
    \centering
        \includegraphics[width=0.45\textwidth]{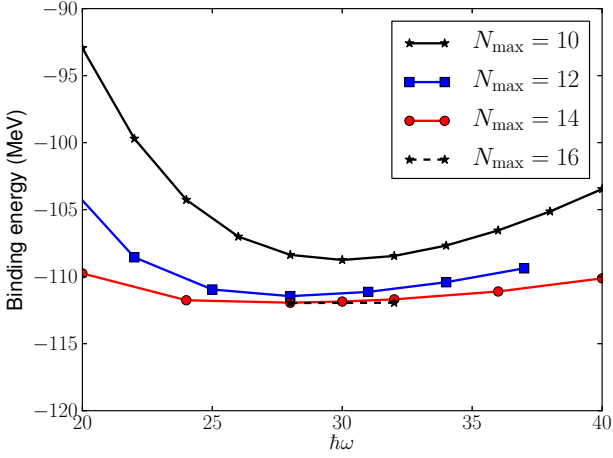}
        \caption{\label{fig:o18_3+1_vbare_j6}(Color online) The total energy of the $J^\pi=3+$
        state in ${}^{18}$O as a function of the oscillator parameter, $\hbar \omega$.
        Different lines correspond to different model spaces, parametrized by the
        variable $\nmax$~\eqref{eq:res_nmax}.}
\end{figure}

Figure~\ref{fig:o18_3+1_vbare_j6} shows the total energy of the first excited
$J^\pi = 3^+$ state in ${}^{18}$O for different model spaces. Here, a shallow
minimum develops at $\hbar \omega = 28$~MeV. Moreover, this state is very well
converged, with a difference in energy of only about $25$~keV between
calculations in the two largest model spaces.
It is clear that the rate of convergence differs for different values of
$\hbar \omega$. When excitation energies are wanted, different choices of $\hbar
\omega$ lead to different results.
Let us discuss two options to evaluate the excitation energy. First, the total energies
can be treated as variational results, where the lowest energy for the
ground state and the lowest energy for the $J^\pi= 3^+$ excited state, are
chosen. Thus, at
$\nmax = 16$ the excitation energy can be calculated as
\begin{equation}
    E_x(3^+) = E_{3^+}(28\mathrm{MeV}) - E_{0_1^+}(32\mathrm{MeV}), \label{eq:res_a18_exenergy_alt}
\end{equation}
where $E_{3^+}(28\mathrm{MeV})$ is the total energy of the
$J^\pi=3^+$ excited state, calculated at $\hbar \omega = 28$~MeV, while
$E_{0_1^+}(32\mathrm{MeV})$ is the ground-state energy calculated
at $\hbar \omega = 32$~MeV.

Second, the same value of
$\hbar \omega$ can be used for both energies, typically where the ground
state has a minimum. Thus, for the current case the excitation energy is
calculated as
\begin{equation}
    E_x(3^+) = E_{3^+}(32\mathrm{MeV}) - E_{0_1^+}(32\mathrm{MeV}). \label{eq:res_a18_exenergy}
\end{equation}
The difference in energy between these two options is minimal if sufficiently
large model spaces are used, but it will have a significant 
impact on the rate of convergence.

\begin{figure}[htp]
    \centering
        \includegraphics[width=0.45\textwidth]{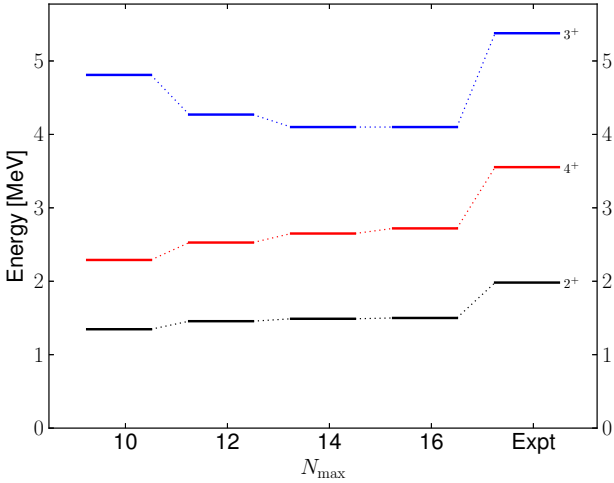}
        \caption{\label{fig:leveltrack_o18_alt}(Color online) The excitation energies of the first
        $J^\pi = 2^+$, $3^+$, and $4^+$ excited states in ${}^{18}$O where the
        best values for $\hbar \omega$ has been chosen for each state. The different  columns represent different model spaces
        parametrized by the variable $\nmax$~\eqref{eq:res_nmax}. The rightmost
        column contains experimental values from \textcite{Tilley1995}.}
\end{figure}

\begin{figure}[htp]
    \centering
        \includegraphics[width=0.45\textwidth]{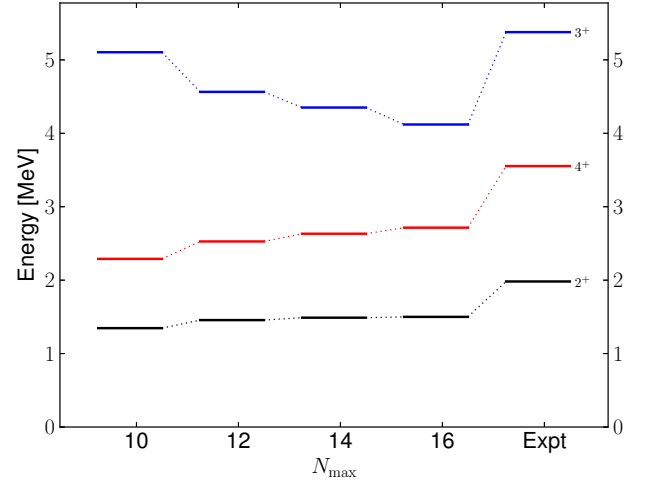}
        \caption{\label{fig:leveltrack_o18}(Color online) The excitation energies of the first
        $J^\pi = 2^+$, $3^+$, and $4^+$ excited states in ${}^{18}$O for $\hbar
        \omega = 32$~MeV. The different  columns represent different model spaces
        parametrized by the variable $\nmax$~\eqref{eq:res_nmax}. The rightmost
        column contains experimental values from \textcite{Tilley1995}.}
\end{figure}

The effect is best viewed in
Figs.~\ref{fig:leveltrack_o18_alt} and \ref{fig:leveltrack_o18}, where the
excitation energies of the first $J^\pi = 2^+$, $3^+$, and $4^+$ excited states in
${}^{18}$O as functions of the size of the model space, are plotted. In
Fig.~\ref{fig:leveltrack_o18_alt}, the excitation energies are calculated according
to Eq.~\eqref{eq:res_a18_exenergy_alt}, while they are calculated according to
Eq.~\eqref{eq:res_a18_exenergy} in Fig.~\ref{fig:leveltrack_o18}.
For ${}^{18}$O, this choice will affect only the $J^\pi=3^+$ state, as the other states
all have  minimum values at $\hbar \omega = 32$~MeV. The effect is significant, but
the first approach of Fig.~\ref{fig:leveltrack_o18_alt} correctly
depicts the level of convergence of the excited states and will be used
in the following.

As in the previous section, let us look at some properties of the wave
functions.
\begingroup
\begin{table*}[htp]
    \begin{ruledtabular}
        \begin{tabular}{ccll}
            State & $n(2p0h)$ & Dominant configuration(s) & Weight(s) \\
            \hline
            ${}^{18}$O($0^+_1$) & 0.87 & $(d)^2$, $(s)^2$ & 0.75,0.12 \\
            ${}^{18}$O($0^+_2$) & 0.87 & $(s)^2$, $(d)^2$ & 0.80, 0.07 \\
            ${}^{18}$O($0^+_3$) & 0.87 & $(d)^2$, $(f)^2$ & 0.85, 0.02  \\
            \\
            ${}^{18}$O($2^+_1$) & 0.88 & $(d)^2$, $(sd)$ & 0.55, 0.32 \\
            ${}^{18}$O($2^+_2$) & 0.88 & $(sd)$, $(d)^2$ & 0.61, 0.27 \\
            ${}^{18}$O($2^+_3$) & 0.88 & $(d)^2$, $(sd)$ & 0.76, 0.11 \\
            \\
            ${}^{18}$O($3^+_1$) & 0.88 & $(sd)$ & 0.88 \\
            ${}^{18}$O($4^+_1$) & 0.88 & $(d)^2$ & 0.88 \\
            \\
            ${}^{18}$O($1^-_1$) & 0.83 & $(dp)$,$(sp)$, $(df)$ & 0.34, 0.26, 0.23 \\
            ${}^{18}$O($2^-_1$) & 0.84 & $(dp)$,$(df)$,$(sp)$ & 0.50, 0.32, 0.02 \\
            ${}^{18}$O($3^-_1$) & 0.82 & $(dp)$,$(df)$,$(sf)$ & 0.42, 0.20, 0.19 \\
                \\
            ${}^{18}$F($0^+_1$) & 0.87 & $(d)^2$, $(s)^2$ & 0.70, 0.16 \\
            ${}^{18}$F($1^+_1$) & 0.89 & $(d)^2$, $(s)^2$,$(f)^2$,$(sd)$ & 0.61,
            0.20, 0.02, 0.01 \\
            ${}^{18}$F($2^+_1$) & 0.88 & $(sd)$, $(d)^2$,$(dg)$,$(f)^2$ & 0.59,
            0.24, 0.02, 0.01 \\
            ${}^{18}$F($3^+_1$) & 0.88 & $(sd)$, $(d)^2$,$(dg)$ & 0.58,
            0.26, 0.02 \\
            ${}^{18}$F($4^+_1$) & 0.88 & $(d)^2$ & 0.88 \\
            ${}^{18}$F($5^+_1$) & 0.88 & $(d)^2$, $(sg)$, $(dg)$ & 0.86, 0.01,
            0.01 \\
        \end{tabular}
    \end{ruledtabular}
    \caption{\label{tab:res_a18_weights} This table shows the $2$p-$0$h partial
    norms\eqref{eq:norm_2p0h}, as well as the dominant configurations for
    calculated states in both ${}^{18}$O and ${}^{18}$F. The weights are
    calculated according to Eq.~\eqref{eq:weights_2p0h} where all nodes for a
    given partial wave contribute to  the sum and spin-orbit partners are not
    distinguished.}
\end{table*}
\endgroup
First, the partial norms defined in Eq.~\eqref{eq:norm_2p0h} and the total
weights~\eqref{eq:weights_2p0h} of the different configurations are calculated. The results are
tabulated in Table~\ref{tab:res_a18_weights}. All positive-parity states have
two nucleons in the $sd$ shell and are consistent with the standard shell model
picture. As in the previous section, all $2$p-$0$h norms are close to $0.90$,
except for the negative-parity states which are closer to $0.80$. The negative-parity states are dominated by cross-shell configurations as these are the only
$2$p-$0$h configurations that can give a negative parity. $3$p-$1$h
excitations from the $p$ shell give a substantial contribution to the $3$p-$1$h
norm.

Second, the center-of-mass contamination of each wave function was analyzed
according to the prescription in Sec.~\ref{sec:com}. Of the states tabulated
in Table~\ref{tab:res_a18_weights}, four states had a significant center-of-mass
contamination. Let us, first, focus on the three negative-parity states in
${}^{18}$O.

\begin{figure}[htp]
    \centering
        \includegraphics[width=0.45\textwidth]{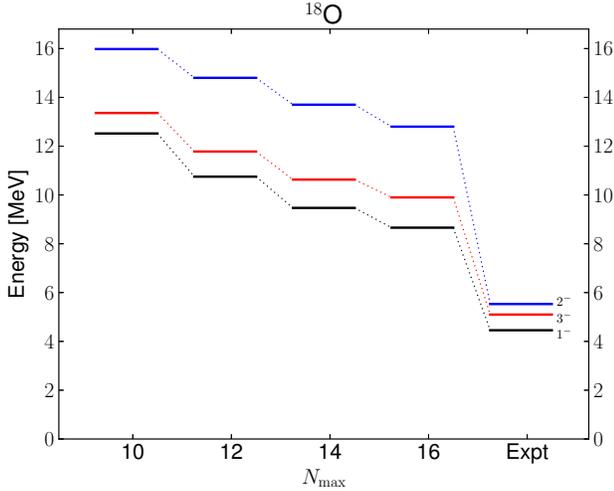}
        \caption{\label{fig:leveltrack_o18_neg}(cClor online) The excitation energies of the first
        $J^\pi = 1^-$, $2^-$, and $3^-$ excited states in ${}^{18}$O for $\hbar
        \omega = 32$~MeV. The different  columns represent different model spaces
        parametrized by the variable $\nmax$~\eqref{eq:res_nmax}. The rightmost
        column contains experimental values from \textcite{Tilley1995}.}
\end{figure}

Figure~\ref{fig:leveltrack_o18_neg} shows the excitation energies of the
negative-parity states in ${}^{18}$O and they are not yet converged at
$\nmax=16$. Although they had a large center-of-mass component, it was not possible to establish what kind of
center-of-mass excitations these states corresponded to. Calculations in larger
model spaces needs to be performed to correctly describe these states. However,
it is also necessary to include $4$p-$2$h correlations to get these states
right.
This can be understood by examining how negative-parity states can occur
in ${}^{18}$O. First, they can be produced by placing one neutron in the $sd$ shell, while
the other is placed in the $pf$ shell. If this was the dominant configuration, the
current truncation would have been enough. Second, they can also be produced by
placing two neutrons in the $sd$ shell and excite a nucleon from the $p$ shell up to
the $sd$ shell. If these kind of excited configurations are comparable in energy
to the first kind, $3$p-$1$h configurations start to dominate and $4$p-$2$h
configurations are necessary for the proper relaxation of the wave function. One
can ask whether the center-of-mass contamination would change if these
configurations were included and whether it is a result of a poorly converged
wave function, but this will be a topic for future work.

In the spectrum of ${}^{18}$O, there are three bound $J^\pi = 0^+$ and $2^+$ states. The
second $J^\pi = 0^+$ state is especially interesting for this method, as it is a
$4$p-$2$h state~\cite{Ellis1970}. In the shell-model language, it is an
intruder state, because configurations outside the $sd$ shell are important to get
this state right. As the current implementation includes only the $3$p-$1$h
configurations, this state can provide clues as to what type of behavior can be
expected from states that are not converged with respect to the level of approximation.

\begin{figure}[htp]
    \centering
        \includegraphics[width=0.45\textwidth]{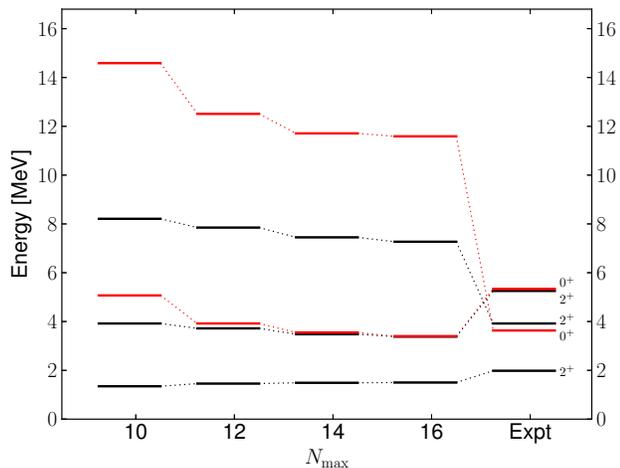}
        \caption{\label{fig:leveltrack_o18_0+}(Color online) The excitation energies of the first
        few $J^\pi = 0+$ and $2+$ states in ${}^{18}$O for $\hbar \omega =
        32$~MeV. The different  columns represents different model spaces
        parametrized by the variable $\nmax$~\eqref{eq:res_nmax}. The rightmost
        column contains experimental values from \textcite{Tilley1995}.}
\end{figure}

Table~\ref{tab:res_a18_weights} lists three $J^\pi=0^+$ states in ${}^{18}$O and
none of them stands out. They all have similar partial norms of around $88$~\%
and are dominated by two neutrons in $d_{5/2}$, $s_{1/2}$, and $d_{3/2}$,
respectively. In Fig.~\ref{fig:leveltrack_o18_0+} the
convergence patterns for these states are plotted, along with those of the three $J^\pi=2^+$ states.
All states show similar level of convergence and it is not possible to single
out one of the states. However, if we look at the center-of-mass contamination of these
states, the third $J^\pi=0^+$ shows a large contamination, while the other
states show almost none. 
Assuming that missing many-body correlations will manifest as larger
center-of-mass contaminations in the final wave functions, this state is associated with the experimental second
$J^\pi=0^+$ state. The calculated second $J^\pi=0^+$ is closer in energy, but
including effects of three-nucleon forces pushes this state higher
in energy and very close to the experimentally observed $J^\pi =0^+$ state at
$5.34$~MeV.~\cite{Hagen2012a}.
A similar effect occurs among the $J^\pi = 2^+$ excited states,
but it is less prominent. Here, the center-of-mass contamination were
negligible for all but the third $J^\pi=2^+$ state, but even here, the
contamination was small compared to the third $J^\pi=0^+$ state.

Let us summarize the discussion of missing many-body correlations. 
Three different markers have been identified to indicate missing physics.
Unfortunately, none of them can be used quantitatively and all must be
evaluated simultaneously to form a general picture.
First, the partial norms can be used to differentiate among different
states. From these calculations it seems that a $2$p-$0$h norm of around $90$~\%
is the standard. A lower partial norm, might indicate the need for $4$p-$2$h or
higher correlations.

Second, we look at the
convergence patterns and if energies converge slowly, this probably means
that something is missing from the calculation. In weakly bound states, for
example, continuum effects result in the need for additional resolution in the
single-particle basis.
Finally, we look at the
level of center-of-mass contamination present in the wave function. Either the
state can be identified as a spurious center-of-mass excitation or a small
non-zero center-of-mass component might indicate missing correlations.
None of these arguments can be analyzed in detail before $4$p-$2$h
configurations are included. This is a work in progress, but,
computationally, it will only be possible to include these configurations in a
small single-particle space. If the $3$p-$1$h and $4$p-$2$h configurations are
defined only in a so-called active space around the Fermi level, the
computational cost might be manageable. This has been done successfully in
\textcite{Gour2005} and should prove to be a valuable approximation also in this
method. The formation of a correlated $\alpha$ cluster around the Fermi level is
important in this mass region  and can
hopefully be accounted for using a minimal set of $4$p-$2$h configurations. 

\begin{figure}[htp]
    \centering
        \includegraphics[width=0.45\textwidth]{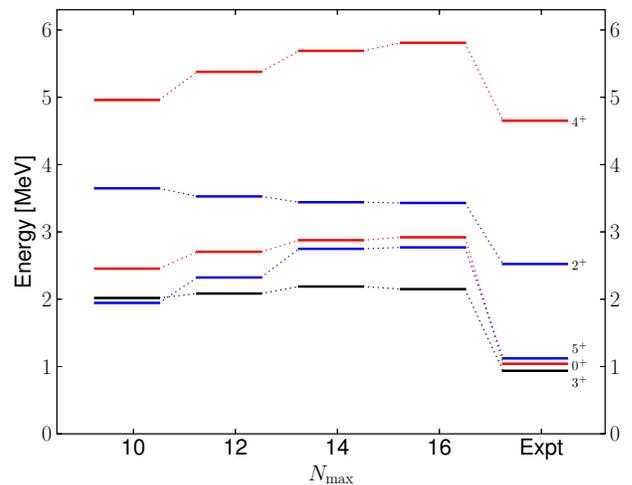}
        \caption{\label{fig:leveltrack_f18}(Color online) The excitation energies of the first
        $J^\pi = 0^+$, $2^+$, $3^+$, $4^+$, and $5^+$ states in ${}^{18}$F,
        calculated at $\hbar \omega=30$~MeV. The different  columns represents different model spaces
        parametrized by the variable $\nmax$~\eqref{eq:res_nmax}. The rightmost 
        column contains experimental values from \textcite{Tilley1995}.}
\end{figure}

Let us also look at the convergence of selected states in ${}^{18}$F.
Figure~\ref{fig:leveltrack_f18} shows the excitation energy of the first few
states in ${}^{18}$F for different model spaces. Here all states are relatively well
converged, with only the $J^\pi = 4^+$ state showing some model space dependence
at $\nmax =16$. None of these states have significant center-of-mass
contamination and all partial norms are on the same level as can be seen in
Table~\ref{tab:res_a18_weights}. This table also shows there is a slight
contribution to the wave function from outside the $sd$ shell.

\begin{figure}[htp]
    \centering
        \includegraphics[width=0.45\textwidth]{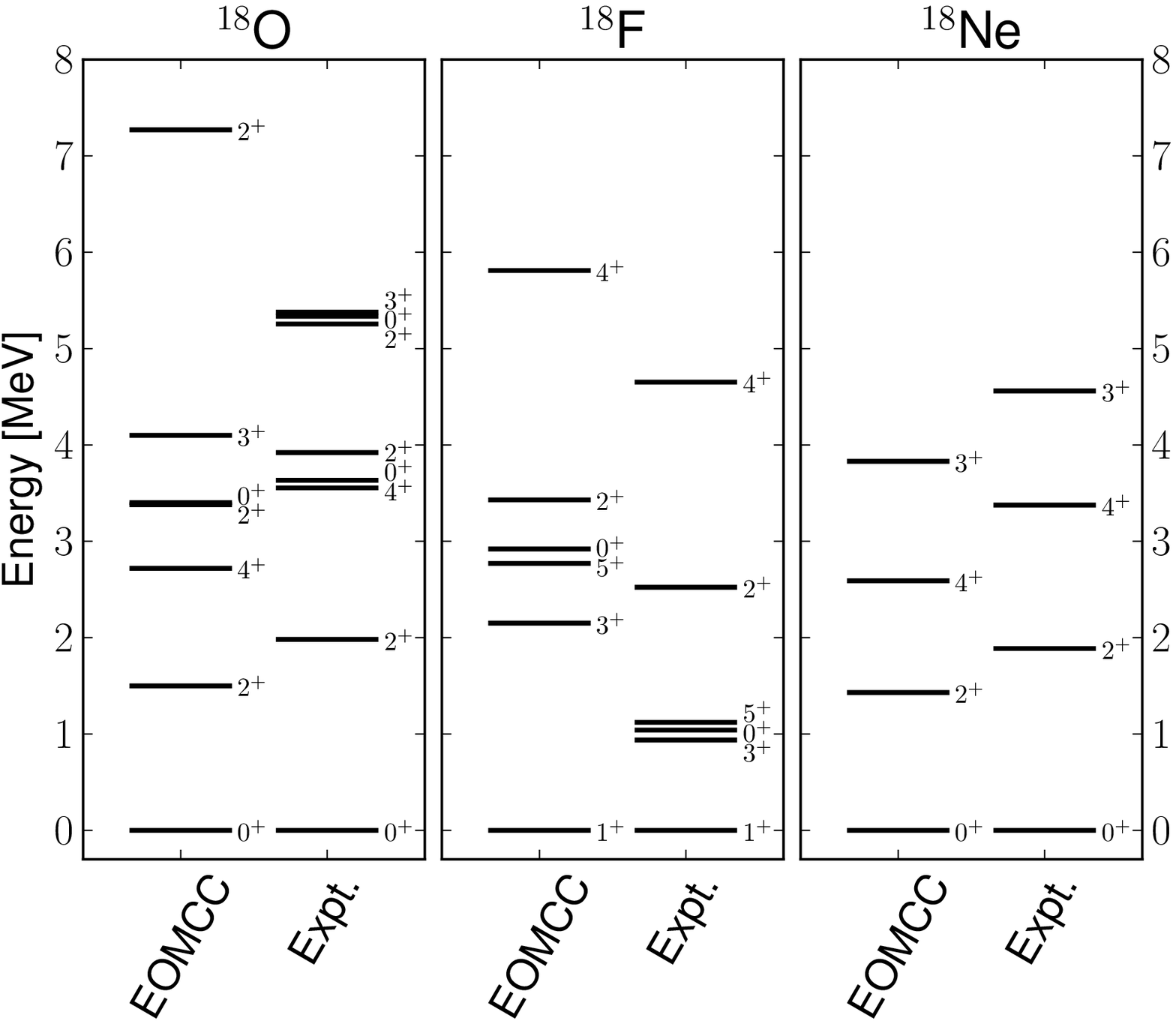}
        \caption{\label{fig:levels_A18}Excitation energies for selected states
        in ${}^{18}$O, ${}^{18}$F, and ${}^{18}$Ne compared to experimental values from
        \textcite{Tilley1995}.}
\end{figure}
Figure~\ref{fig:levels_A18} shows the excitation spectra of ${}^{18}$O, ${}^{18}$F, and
${}^{18}$Ne. Only states that are considered good are plotted and compared to
data.

\begin{table}
    \begin{ruledtabular}
    \begin{tabular}{cccc}
    \multirow{2}{*}{${}^{18}$O} & \multirow{2}{*}{Expt.} & \nlo \\
    & & ($\Lambda=500\mathrm{MeV}$) \\
    \hline
    $\mathrm{E_{gs}}(0^+)$ & $-139.807$ & $ -130.00(2) $ \\
    \hline
    $\mathrm{E_{gs}}(0^+_1)$ & $0.0$ & $0.0$ \\
    $\mathrm{E_x}(2^+_1)$ & $+1.982$ & $+1.50(1)$  \\
    $\mathrm{E_x}(4^+_1)$ & $+3.554$ & $+2.72(7)$  \\
    $\mathrm{E_x}(0^+_2)$ & $+3.633$ & NA \\
    $\mathrm{E_x}(2^+_2)$ & $+3.92$ & $+3.38(2)$  \\
    $\mathrm{E_x}(2^+_3)$ & $+5.255$ & $+7.27(18)$  \\
    $\mathrm{E_x}(0^+_3)$ & $+5.336$ & $+3.40(15)$ \\
    $\mathrm{E_x}(3^+_1)$ & $+5.3778$ & $+4.12(1)$ \\
    \hline
    \multirow{2}{*}{${}^{18}$F} \\
    \\
    \hline
    $\mathrm{E_{gs}}(1^+)$ & $-137.370$ & $ -129.75(16) $\\
    \hline 
    $\mathrm{E_{gs}}(1^+)$ & $0.0$ & $0.0$\\
    $\mathrm{E_x}(3^+)$ & $+0.937$ & $ +2.15(4)$   \\
    $\mathrm{E_{x}}(0^+)$ & $+1.041$ & $ +2.92(4)$ \\
    $\mathrm{E_x}(5^+)$ & $+1.121$ & $+2.77(2)$    \\
    $\mathrm{E_x}(2^+)$ & $+2.523$ & $+3.43(1)$    \\
    $\mathrm{E_x}(4^+)$ & $+4.652$ & $+5.81(12)$   \\
    \hline
    \multirow{2}{*}{${}^{18}$Ne} \\
    \\
    \hline
    $\mathrm{E_{gs}}(0^+)$ &  $-132.143$ & $ -122.56(15) $ \\
    \hline
    $\mathrm{E_{gs}}(0^+)$ & $0.0$ & $0.0$ \\
    $\mathrm{E_x}(2^+)$ &  $+1.887$ & $ +1.43(3)$ \\
    $\mathrm{E_x}(4^+)$ & $+3.376$ & $ +2.59(6)$ \\
    $\mathrm{E_x}(3^+)$ & $+4.561$ & $+3.83(4)$ \\
\end{tabular}

    \end{ruledtabular}
    \caption{\label{tab:result_A18}Ground- and excited-state energies for the $A=18$ system, compared to experimental data. The experimental energies are from \textcite{Tilley1995}. The number in parenthesis indicates the level of convergence and is the difference in energy between calculations in the two largest model spaces.}
\end{table}

For future comparison, Table~\ref{tab:result_A18} lists the numerical values used in
Fig.~\ref{fig:levels_A18}, together with the ground-state energies. The
uncertainty is calculated as the difference in energy between calculations in
the two largest model spaces. The experimental values are from
\textcite{Tilley1995}.

The total binding energy of ${}^{18}$O is comparable to what was found in \textcite{Hergert2013b},
where the in-medium similarity renormalization
group~(IM-SRG)~\cite{Tsukiyama2011} method was used to
compute the ground-state energies of even oxygen isotopes. Although they used an
SRG evolved interaction based on chiral interaction at fourth order by
\textcite{Entem2003}, induced three-nucleon forces were also included in
the final calculation to make the results comparable to those in
Table~\ref{tab:result_A18}.

Compared to data, the level ordering in ${}^{18}$F is reproduced, but the
excitation energies are systematically overestimated. Disregarding the missing states
in ${}^{18}$O, the level ordering is also reproduced, but here the excitation
energies are systematically underestimated. This is consistent with shell-model
calculations~\cite{Holt2005,Dong2011} of these nuclei using different model-space interactions. For results based on the chiral interactions used in this
work, the inclusion of three-nucleon forces~\cite{Hagen2012a,Holt2011} gives
results that better match experimental data. Recently,
\textcite{Ekstrom2013} showed that the effects of three-nucleon forces
depends on the low-energy constants used in the parametrization of the
chiral potential. To accurately evaluate the quality of these forces will
require not only three-nucleon forces and continuum degrees of freedom but also
additional correlations in the many-body
wave function~\cite{Volya2005,Hagen2009,Otsuka2010,Hagen2012a,Hergert2013,Hergert2013b}. 

\section{\label{sec:conclusion}Conclusions and outlook}
The spherical version of the 2PA-EOM-CCSD method has been presented. This is appropriate
for the calculation of energy eigenstates in nuclei that can be described as two
particles attached to a closed (sub-)shell reference. The method has been
evaluated in both $A=6$ and $A=18$ nuclei, where the results were converged with
respect to the single-particle basis.

It was also shown that the wave function from a 2PA-EOM-CCSD calculation
separates into an intrinsic part and a Gaussian for the center-of-mass
coordinate, not necessarily the ground state of the harmonic oscillator
Hamiltonian. Wave functions with significant center-of-mass contamination were
either identified as a spurious center-of-mass excitation or were not converged
with respect to the current approximation level.

In comparison with a full diagonalization, both ground-state and excited-state energies were in general very accurate. However, one excited state in ${}^{6}$Li
deviated significantly from the ``exact'' result, showing the need to include
additional correlations like $4$p-$2$h configurations for the accurate treatment
of complex states. For
simple states, where a $2$p structure is dominant, the current level of
truncation is adequate.

Both three-nucleon forces and a correct treatment
of the scattering continuum are needed to refine the results.

\begin{acknowledgments}
I thank M.~Hjorth-Jensen and T.~Papenbrock for valuable comments
on the manuscript. In addition I thank G.~Hagen
and A.~Ekstr\"om for very useful discussions. 
This work was partly supported by the Office of Nuclear Physics, U.S.
Department of Energy (Oak Ridge National Laboratory), under Contracts No.
DE-FG02-96ER40963 (University of Tennessee) and No.DE-SC0008499 (NUCLEI
SciDAC-3 Collaboration).
An award of computer time was provided by the Innovative and Novel
Computational Impact on Theory and Experiment (INCITE) program.  This
research used resources of the Oak Ridge Leadership Computing Facility
located in the Oak Ridge National Laboratory, which is supported by the
Office of Science of the Department of Energy under Contract No.
DE-AC05-00OR22725 and used computational resources of the
National Center for Computational Sciences, the National Institute for
Computational Sciences, and the Notur project in Norway.
\end{acknowledgments}

\appendix

\section{\label{app:reduced}Reduced matrix elements}
The reduced matrix elements of a spherical tensor operator $\op{T}_M^J$ of rank
$J$ and projection $M$, are defined according to the
\WE{},
\begin{equation}
    \bra{\alpha; J_\alpha M_\alpha}\op{T}^J_M \ket{\beta; J_\beta M_\beta} = 
    C^{ J J_\beta J_\alpha }_{M M_\beta M_\alpha}
    \bra{\alpha; J_\alpha}| \op{T}^J |\ket{\beta; J_\beta}.
    \label{eq:reduced_matrix_element}
\end{equation}
Here $\alpha$ and $\beta$ are general labels representing all quantum numbers
except angular momentum and its projection, while $J_\alpha(M_\alpha)$ and
$J_\beta(M_\beta)$ are the total angular momentum(projection) of the bra and ket
states, respectively. The double bars denote reduced matrix elements and does not depend on
any of the angular-momentum projections and $C^{ J J_\beta J_\alpha }_{M M_\beta
M_\alpha}$ is a \CG.

In coupled cluster, the unknown amplitudes are the matrix elements of the
cluster operator $\op{T}^0_0$,
\begin{align}
    t_i^a &= \bra{a; j_a m_a}\op{T}^0_0 \ket{i; j_i m_i} \\
    t_{ij}^{ab} &= \bra{ab; j_a m_a j_b m_b} \op{T}^0_0 \ket{ij; j_i m_i j_j
    m_j} \\
    \vdots &= \vdots,
\end{align}
where the operator sub- and superscriptsidentify the cluster operator as a
scalar under rotation. Also, the labels $abij\ldots$, denote single-particle
states and we have singled out the angular momentum(projection) in the labels
$j_a(m_a)$, and so on.

Now the reduced matrix elements of the cluster operator are defined according
to Eq.~\eqref{eq:reduced_matrix_element}. For example, 
\begin{multline}
    \bra{ab; j_a m_a j_b m_b} \op{T}^0_0 \ket{ij; j_i m_i j_jm_j} = \\
    \sum_{\substack{
    J_{ab} M_{ab} \\
    J_{ij} M_{ij}}}
    C_{m_a m_b M_{ab}}^{j_a j_b J_{ab}} C_{m_i m_j M_{ij}}^{j_i j_j J_{ij}}
    C_{0 M_{ij} M_{ab}}^{0 J_{ij} J_{ab}} \\
    \times \bra{ab; j_a j_b; J_{ab}}| \op{T}^0 |\ket{ij; j_i j_j; J_{ij}},
\end{multline}
where the sum and the first two \CG s come from the coupling of $j_a$ and $j_b$
to $J_{ab}$ and of $j_i$ and $j_j$ to $J_{ij}$, in that specific order This expression is simplified by
the explicit evaluation of the third \CG
\begin{equation}
    C_{0 M_{ij} M_{ab}}^{0 J_{ij} J_{ab}} = \delta_{J_{ab}, J_{ij}=J}
    \delta_{M_{ab}, M_{ij}=M},
\end{equation}
where $\delta_{\alpha, \beta}$ is the Kronecker $\delta$. We get
\begin{multline}
    \bra{ab; j_a m_a j_b m_b} \op{T}^0_0 \ket{ij; j_i m_i j_jm_j} = \\
    \sum_{J_{} M_{}}
    C_{m_a m_b M}^{j_a j_b J} C_{m_i m_j M}^{j_i j_j J}
    \bra{ab; j_a j_b; J}| \op{T}^0 |\ket{ij; j_i j_j; J}.
\end{multline}
When this specific coupling order is used~(left to right) and no confusion will
arise, we will use a shorthand notation for the reduced matrix elements, defined by
\begin{align}
    t_i^a(J) &= \bra{a; J}|\op{T}^0 |\ket{i; J} \\
    t_{ij}^{ab}(J) &= \bra{ab; j_a j_b; J}| \op{T}^0 |\ket{ij; j_i j_j; J}.
\end{align}

The transformations
between the reduced amplitudes
and the original amplitudes of the cluster operator are defined as
\begin{align}
    t_i^a(J) &= \delta_{j_a, j_i=J} t_i^a \label{eq:reduced_t1}\\
    t_i^a &= \delta_{j_a, j_i=J} t_i^a(J) \label{eq:reduced_t1_rev}\\
    t_{ij}^{ab}(J) &=  \frac{1}{\hat{J}^2}
    \sum_{\substack{
        m_a m_b \\
        m_i m_j M}}
        C_{m_a m_b M}^{j_a j_b J} C_{m_i m_j M}^{j_i j_j J}
        t_{ij}^{ab} \label{eq:reduced_t2} \\
    t_{ij}^{ab} &= 
        \sum_{J M}
    C_{m_a m_b M}^{j_a j_b J} C_{m_i m_j M}^{j_i j_j J}
    t_{ij}^{ab}(J) \label{eq:reduced_t2_rev},
\end{align}
where we use the convention $\hat{J} = \sqrt{2J+1}$. 

As the similarity-transformed Hamiltonian($\barh$) is a scalar under rotation as well,
the shorthand form of its reduced matrix elements are defined analogously,
\begin{align}
    \barh_q^p(J) &= \bra{p; J}|\barh^0 |\ket{q; J} \\
    \barh_{rs}^{pq}(J) &= \bra{pq; j_p j_q; J}| \barh^0 |\ket{rs; j_r j_s; J}.
\end{align}
The transformations between the original and reduced matrix elements are given
by
\begin{align}
    \barh_q^p(J) &= \delta_{j_p, j_q=J} \barh_q^p \label{eq:reduced_barh1}\\
    \barh_q^p &= \delta_{j_p, j_q=J} \barh_q^p(J) \label{eq:reduced_barh1_rev}\\
    \barh_{rs}^{pq}(J) &=  \frac{1}{\hat{J}^2}
    \sum_{\substack{
        m_p m_q \\
        m_r m_s M}}
        C_{m_p m_q M}^{s_p s_q J} C_{m_r m_s M}^{s_r s_s J}
        \barh_{rs}^{pq} \label{eq:reduced_barh2} \\
    \barh_{rs}^{pq} &=
        \sum_{J M}
    C_{m_p m_q M}^{s_p s_q J} C_{m_r m_s M}^{s_r s_s J}
    \barh_{rs}^{pq}(J) \label{eq:reduced_barh2_rev}.
\end{align}
The original matrix elements of the similarity-transformed Hamiltonian are defined in \textcite{Jansen2011}.

The $r$ amplitudes are the matrix elements of the excitation operator in
Eq.~\eqref{eq:eom_operator_sph},
\begin{align}
    r^{ab} &= \bra{a b; j_a m_a j_b m_b} \op{R}_M^J \ket{0} \\
    r^{abc}_i &= \bra{a b c; j_a m_a j_b m_b j_c m_c} \op{R}_M^J \ket{i; j_i m_i},
\end{align}
where the operator is now a general tensor operator of rank $J$ and projection
$M$. The bra side contains up to three indices and we have to couple three angular
momentum vectors to be able to define the reduced amplitudes. Using
$r^{abc}_i$ as an example, we couple from left to right and get
\begin{multline}
    \bra{a b c; j_a m_a j_b m_b j_c m_c} \op{R}_M^J \ket{i; j_i m_i} = \\
    \sum_{\substack{
    J_{ab} M_{ab} \\
    J_{abc} M_{abc}}}
    C_{m_a m_b M_{ab}}^{j_a j_b J_{ab}} C_{M_{ab} m_c M_{abc}}^{J_{ab} j_c
    J_{abc}} C_{M m_i M_{abc}}^{J j_i J_{abc}} \\
    \bra{a b c; j_a j_b; J_{ab} j_c; J_{abc}}|\op{R}^J|\ket{i; j_i},
\end{multline}
where the last \CG{} is due to the \WE{}.
We let the order of the angular-momentum labels on the bra side specify the
coupling order, where $j_a$ and $j_b$ couples to $J_{ab}$. In turn,
$J_{ab}$ and $j_c$ couples to $J_{abc}$. When this coupling order has been used
and no confusion will arise, we will use the shorthand notation for the reduced
elements, defined by
\begin{align}
    r^{ab}(J) &= \bra{ab; j_a j_b; J}|\op{R}^J|\ket{o} \\
    r^{abc}_i\left(J, J_{abc}, J_{ab}\right) &= \bra{a b c; j_a j_b; J_{ab} j_c; J_{abc}}|\op{R}^J|\ket{i;
    j_i}.
\end{align}

In the shorthand notation, the transformations between the reduced and the
original amplitudes is
\begin{align}
    r^{ab}(J) &= 
        \frac{1}{\hat{J}^2} 
        \sum_{M m_a m_b}
        r^{ab}
        C^{j_a j_b J}_{m_a m_b M}
        \label{eq:reduced_2p} \\
    r^{ab} &= 
        C^{j_a j_b J}_{m_am_b M}
        r^{ab}(J)
        \label{eq:reduced_2p_rev}
        \\
    \begin{split}
    r^{abc}_{i}(J,J_{abc}, J_{ab}) &= 
        \frac{1}{\hat{J}^2_{abc}} 
        \sum_{\substack{
            M M_{abc} M_{ab} \\
            m_a m_b m_c m_i 
        }}
        r^{abc}_i
        C^{j_a j_b J_{ab}}_{m_a m_b M_{ab}}
        \\
        & \quad 
        \times
        C^{J_{ab} j_c J_{abc}}_{M_{ab} m_c M_{abc}}
        C^{J j_i J_{abc}}_{M m_i M_{abc}}.
    \end{split}
        \label{eq:reduced_3p1h} \\
    \begin{split}
    r^{abc}_{i} &= 
        \sum_{\substack{
        J_{abc} M_{abc} \\
        J_{ab} M_{ab}
        }}
        r^{abc}_i(J, J_{abc}, J_{ab}) \\
        & \quad 
        \times
        C^{j_a j_b M_{ab}}_{m_a m_b M_{ab}}
        C^{J_{ab} j_c J_{abc}}_{M_{ab} m_c M_{abc}}
        C^{J j_i J_{abc}}_{M m_i M_{abc}}.
    \end{split}
        \label{eq:reduced_3p1h_rev}
\end{align}

\section{\label{app:permutation}Permutation operators}
The diagrams in Table~\ref{tab:2pa_all} contain permutation operators that
guarantees an antisymmetric final wave function. In the uncoupled formalism,
these were simple and defined by
\begin{align}
    \op{P}(ab) &= \op{1} - \op{P}_{a,b} \label{perm_op_2p_old}\\
    \op{P}(ab,c) &= \op{1} - \op{P}_{a,c} - \op{P}_{b,c},
    \label{eq:perm_op_3p1h_old}
\end{align}
where $\op{P}_{a,b}$ permutes indices $a$ and $b$ and $\op{1}$ is the identity
operator.
In the spherical formalism, this simple form is not adequate, as a specific
coupling order is used in all reduced amplitudes. To see why, let us apply
$\op{P}_{a,b}$ to a reduced amplitude $r^{ab}(J)$
\begin{equation}
    \op{P}_{a,b} \bra{ab; j_a j_b; J}|\op{R}^J|\ket{0}  = \bra{ba; j_a j_b;
    J}|\op{R}^J|\ket{0}
\end{equation}
While this coupling order is the correct order when calculating the contribution
to $\left(\barh \op{R} \right)^{ab}$, the reduced amplitudes are defined in a
different coupling order. To compensate, we change the coupling order and
introduce a phase~(see \textcite{edmunds1960} for details),
\begin{equation}
    \ket{ba; j_a j_b; J} =
    (-1)^{j_a + j_b -J} \ket{ba; j_b j_a; J}.
\end{equation}
Thus, we define
\begin{align}
    \op{P}(ab) &= \op{1} - (-1)^{j_a + j_b -J}\op{P}_{a,b}, \label{eq:perm_op_2p}
\end{align}
and applied to the reduced amplitude $r^{ab}(J)$ this has the correct form
\begin{equation}
    \op{P}(ab) r^{ab}(J) = r^{ab}(J) - (-1)^{j_a + j_b -J} r^{ba}(J).
\end{equation}

The permutation operators $\op{P}_{a,c}$ and $\op{P}_{b,c}$ in
Eq.~\eqref{eq:perm_op_3p1h_old} are a bit more complicated, as they involve
three particle states. We use standard expressions for coupling three angular
momenta
\begin{align}
    \ket{cba; j_a j_b; J_{ab} j_c; J_{abc} M_{abc}} &= 
        -\sum_{J_{cb}} \hat{J}_{cb} \hat{J}_{ab}
        \begin{Bmatrix}
            j_c & j_b & J_{cb} \\
            j_a & J_{abc} & J_{ab}
        \end{Bmatrix} \nn
    & \quad \times
    \ket{cba; j_c j_b; J_{cb} j_a; J_{abc} M_{abc}} \\
    \ket{acb; j_a j_b; J_{ab} j_c; J_{abc} M_{abc}}
        &=
        \sum_{J_{ac}}
        (-1)^{j_b + j_c -J_{ab} + J_{ac}}
        \hat{J}_{ab} \hat{J}_{ac} \nn
        & \quad \times
        \begin{Bmatrix}
            j_c & j_a & J_{ac} \\
            j_b & J_{abc} & J_{ab}
        \end{Bmatrix} \nn
    & \quad \times
    \ket{acb; j_a j_c; J_{ac} j_b; J_{abc} M_{abc}}
\end{align}

Thus, we define
\begin{multline}
    \op{P}(ab,c) = \op{1} + 
        \sum_{J_{cb}} \hat{J}_{cb} \hat{J}_{ab}
        \begin{Bmatrix}
            j_c & j_b & J_{cb} \\
            j_a & J_{abc} & J_{ab}
        \end{Bmatrix}
        \op{P}_{a,c} - \\
        \sum_{J_{ac}}
        (-1)^{j_b + j_c -J_{ab} + J_{ac}}
        \hat{J}_{ab} \hat{J}_{ac}
         \times
        \begin{Bmatrix}
            j_c & j_a & J_{ac} \\
            j_b & J_{abc} & J_{ab}
        \end{Bmatrix}
        \op{P}_{b,c},
    \label{eq:perm_op_3p1h}
\end{multline}
Since this anti symmetrization contributes a significant part of the overall
calculation, all diagrams containing this operator are applied only once to the
sum of all diagrams containing this operator.

\section{\label{app:threebody}Three-body parts of $\barh$}
There are two three-body matrix elements of $\barh$ that contribute to the
$3$p-$1$h amplitudes. But since the original Hamiltonian does not contain
three-body elements, these deserve special attention. These elements can be
factorized in just the same way as the coupled-cluster amplitude equations, to reduce the
computational cost of these diagrams.  The two three-body contributions to the
$3$p-$1$h amplitudes are
\begin{align}
    (\barh \op{R})^{abc}_i & \gets
    \frac{1}{2}
    \barh_{efi}^{abc} 
    r^{ef}
    +
    \frac{1}{2}
    P(a, bc) \barh_{efi}^{bmc}
    r^{aef}_m \nn
    &=
    -
    \frac{1}{2}
    P(a,bc)
    \barh_{ef}^{am}
    r^{ef}
    t_{mi}^{bc}
    +
    \frac{1}{2}
    P(a,bc)
    \barh_{ef}^{mn}
    r^{ef}
    t_{ni}^{bc}
    t_m^a \nn
    & \qquad
    +
    \frac{1}{2}
    P(a,bc)
    \barh_{ef}^{mn}
    t_{ni}^{bc}
    r_m^{aef}.
\end{align}
These terms are factorized to get
\begin{equation}
    (\barh \op{R})^{abc}_i \gets
    \frac{1}{2}
    P(a,bc)
    t_{im}^{ab} \chi_{m}^{c}
\end{equation}
where we have defined the intermediate
\begin{equation}
    \chi_{m}^{c} = 
    \barh_{ef}^{cm} r^{ef}
    +
    \barh_{ef}^{mn}
    t_n^c
    +
    r_n^{efc}. \label{eq:app_three_intermediate}
\end{equation}
Note that we have swapped the indices to facilitate the angular-momentum
coupling. Now the angular-momentum coupling in the coupled-cluster amplitudes match the coupling in the $3$p-$1$h
amplitudes, so we do not need to break these couplings when rewriting the
diagram in a spherical basis.

In the spherical formulation, it is clear that $\chi_m^c$ are the reduced
matrix elements of the tensor operator $\op{\chi}^J$, which has the same rank
as $\op{R}^J$.  This is a consequence of the scalar character of $\barh$.  By
coupling the matrix elements in Eq.~\eqref{eq:app_three_intermediate} to form reduced matrix elements, we get
the following expression for the reduced matrix elements of $\op{\chi}^J$:
\begin{multline}
    \chi_m^c(J)
    =
    -r^{ef}(J)
    \barh_{ef}^{mc}(J)
    +
    r^{ef}(J)
    \barh_{ef}^{mn} (J)
    t_n^c \\
    +
    \sum_{J_{ef}, J_{efc}}
    (-1)^{j_c + j_m + J_{ef} - J}
    \frac{\hat{J} \hat{J}_{ef}}{\hat{J}^2_{efc}}
    \begin{Bmatrix}
        j_c & j_m & J \\
        j_m & J_{efc} & J_{ef}
    \end{Bmatrix} \\
    \times
    r_n^{efc} (J_{ef}, J_{efc}, J)
    \barh_{ef}^{mn}(J_{ef}).
\end{multline}
    
\end{document}